%% file: coherence-v30.tex
\documentclass[aps,prd,twocolumn,showpacs,amsmath,amssymb,superscriptaddress,nofootinbib,floatfix,10pt,preprintnumbers]{revtex4-1}

\pdfoutput=1
\usepackage[usenames,dvipsnames]{color}
\include{jdefs}
\usepackage{graphicx}
\usepackage{multirow}
\usepackage{graphicx}
\usepackage{url}
\usepackage{xspace}
\usepackage{braket}
\usepackage{tabularx}
\definecolor{darkblue}{rgb}{0,0,0.5}
\definecolor{darkgreen}{rgb}{0.1,0,0.3}
\definecolor{darkred}{rgb}{0.6,0,0}
\usepackage[bookmarks=true, bookmarksnumbered=true, colorlinks=true,
  urlcolor=darkblue, citecolor=darkgreen, linkcolor=darkred,
  breaklinks=true]{hyperref} 
\usepackage[normalem]{ulem}

\usepackage{fullpage}
\usepackage{geometry}      
\newcommand{\st}{\sin^2 \! \theta_W}
\newcommand{\phiB}{\phi_\nu^{^8\mathrm{B}}}
\newcommand{\phipp}{\phi_\nu^{pp}}

\newcommand{\druty}{{\rm events\,ton^{-1}\,yr^{-1}\,keV^{-1}}}

\begin{document}

\title{Physics from solar neutrinos in dark matter direct detection experiments}
 \author{David G. Cerde\~no}
 \affiliation{Institute for Particle Physics Phenomenology (IPPP), Durham University, Durham DH1 3LE, UK}
 \author{Malcolm Fairbairn}
\affiliation{Physics, King's College London, Strand, London, WC2R 2LS, UK}
 \author{Thomas Jubb}
 \affiliation{Institute for Particle Physics Phenomenology (IPPP), Durham University, Durham DH1 3LE, UK}
\author{Pedro~A.~N.~Machado}
\affiliation{Departamento de F\'isica Te\'orica, Universidad Aut\'onoma de Madrid, Cantoblanco E-28049 Madrid, Spain}
\affiliation{Instituto de F\'isica Te\'orica UAM/CSIC,
 Calle Nicol\'as Cabrera 13-15, Cantoblanco E-28049 Madrid, Spain}
\author{Aaron C. Vincent}
 \affiliation{Institute for Particle Physics Phenomenology (IPPP), Durham University, Durham DH1 3LE, UK}
 \author{C\'eline B\oe hm}
 \affiliation{Institute for Particle Physics Phenomenology (IPPP), Durham University, Durham DH1 3LE, UK}
\affiliation{LAPTH, U.~de Savoie, CNRS,  BP 110, 74941 Annecy-Le-Vieux, France}

 \begin{abstract}
The next generation of dark matter direct detection experiments will
be sensitive to both coherent neutrino-nucleus and neutrino-electron
scattering.  This will enable them to explore aspects of solar
physics, perform the lowest energy measurement of the weak angle
  $\sin^2\!\theta_W$ to date, and probe contributions from new
theories with light mediators.  In this article, we
compute the projected nuclear and electron recoil rates expected in
several dark matter direct detection experiments due to solar
neutrinos, and use these estimates to quantify errors on future
measurements of the neutrino fluxes, weak mixing angle and solar
observables, as well as to constrain new physics in the neutrino
sector. Our analysis shows that the combined rates of solar
neutrino events in second generation experiments (SuperCDMS and LZ)
can yield a measurement of the $pp$ flux to 2.5\% accuracy via
electron recoil, and slightly improve the $^8$B flux
determination. Assuming a low-mass argon phase, projected tonne-scale
experiments like DARWIN can reduce the uncertainty on both the $pp$
and boron-8 neutrino fluxes to below 1\%. Finally, we use current
results from LUX, SuperCDMS and CDMSlite to set bounds on new
interactions between neutrinos and electrons or nuclei, and show that
future direct detection experiments can be used to set complementary
constraints on the parameter space associated with light mediators.
\end{abstract}

\maketitle


\section{Introduction}
\label{sec:intro}
Direct detection (DD) experiments are the primary tool in the search
for weakly-interacting dark matter.  Such detectors are sensitive to
nuclear recoil signals in the $\sim 1-100$~keV range from the
scattering of dark matter particles as they stream through the Earth.
These experiments typically consist of dense crystals (such as
germanium, silicon, or sodium iodide) or liquid noble gases such as
xenon or argon. They are located deep in underground mines, shielded
from cosmic rays and cosmogenic radiation. DD experiments thus share
many similarities with underground neutrino detectors and, as such,
may be used to measure neutrino properties. Moreover, their low
  recoil energy threshold makes them excellent tools to search for
  new physics at low scales.

In particular, the next generation (G2) of DD experiments is expected to 
detect neutrinos from nuclear reactions occurring inside the Sun. Future experiments might  detect atmospheric neutrinos 
induced by the interactions of the cosmic rays with the atmosphere, as well as the diffuse background of neutrinos produced in type II supernovae throughout the history of the Universe.  
 
 The coherent  scattering of these neutrinos with nuclei in direct detection experiments constitutes a severe limitation to the detection of dark matter, since their recoil energy is expected to be similar. 
This is referred to as the ``neutrino scattering floor''~\cite{Cabrera:1984rr,Drukier:1986tm,Monroe:2007xp,Vergados:2008jp,Strigari:2009bq,Gutlein:2010tq}. 

For example, the recoil spectrum of a 6 GeV dark matter particle would
be very difficult to distinguish from the $^8$B solar neutrino
flux,  though one may be able to   discriminate both signals by exploiting their
different contributions to annual modulation~\cite{Billard:2013qya,Davis:2014ama}, or 
by using a combination of complementary targets~\cite{Ruppin:2014bra} and  directional
detectors~\cite{Grothaus:2014hja,O'Hare:2015mda} or detectors with improved energy resolution \cite{Dent:2016iht}.

The detection of coherent neutrino scattering is interesting in itself
as this Standard Model (SM) prediction has never been observed in
dedicated neutrino experiments, due to the small cross sections and
the very low recoil energy involved. Moreover, any deviation to the SM
expectation could indicate the existence of new physics at low energy,
while the lack of deviation could help to set constraints on light
mediators.  This attractive possibility is already an integral part of
the science programme of direct detection collaborations, and
presumably within reach of the future SuperCDMS SNOLAB phase
\cite{Brink:2012zza} and LZ \cite{Akerib:2015cja} experiments.

In this work, we focus on the possibility of characterising the
physics of solar neutrinos,  and of probing
new physics with future dark matter experiments. We quantify the
  precision with which  $\st$ can be measured at the lowest possible energy
  scale to date. This can both confirm a long-standing prediction of the standard model; and help search for or rule out effects of new physics, such as a light dark sector, which could change the
  running of $\st$ at low energies (see
  e.g. Ref.~\cite{Davoudiasl:2014kua}).

Before turning to our results, we begin by briefly summarizing the basics of
solar neutrino physics and neutrino scattering off nuclei and
electrons in Sections~\ref{sec:solarneutrino} and \,\ref{sec:CNS},
respectively. In Section\,\ref{sec:solar}, we explore the
reconstruction of solar parameters from the combination of data from
various DD experiments. In Section\,\ref{sec:newphysics}, we determine
the constraints that future DD experiments will be able to place on
new physics models with light mediators. For concreteness, we take a
simplified effective model approach (below the electroweak scale), and
constrain the mediator mass and couplings to electrons, quarks, and
neutrinos.  Finally, we illustrate our results in the case of a light
$U(1)_{B-L}$ gauge boson, showing that current experiments (SuperCDMS,
CDMSlite and LUX) can exclude new regions of the parameter space.  Our
conclusions are presented in Section\,\ref{sec:conclusions}.

\section{Solar neutrinos  \label{sec:solarneutrino}}
In this section and the next, we review the necessary physics of solar neutrino fluxes and direct detection experiments
   that are relevant for this study and that go into the production of our results.

\subsection{Neutrino contribution} 
The dominant contributions to the
neutrino flux in the lowest energy range arise from the various nuclear fusion and decay
processes occurring in the solar core, associated with the Sun's
energy production. The primary fusion process in the Sun is $p + p \rightarrow
^2\!\!\mathrm{H} + e^+ + \nu_e$ and leads to the production of  neutrinos in a continuum up to $E_\nu \lesssim 400$ keV. These are referred to as 
$pp$  neutrinos and are by far the largest contributors to the solar neutrino flux below the MeV scale. Recoils from
$pp$ neutrinos scattering on target nuclei are virtually undetectable,
since the typical momentum transfer is much lower than the threshold
energy of a few keV of current DD experiments. However, the electron recoil energies fit
comfortably above the $\sim$ keV threshold. 

At higher energies but with lower flux (by around three orders of magnitude) we find the  neutrinos produced from the CNO cycles, which we will refer to simply as the CNO neutrinos.  Within the same energy range there are the monoenergetic neutrino lines at $E_\nu = 862$~keV and $384$~keV from $^7$Be.  These energies are typically too low to give rise to a nuclear recoil within the range 1-100 keV that 
current DD experiments are typically optimised for, and too high to give an electron recoil in the right energy range. Some high-metallicity solar models \cite{Vagnozzi:2016cmr} predict CNO fluxes that are over 50\% larger than the expected values in standard solar models. However, these also yield other fluxes that are experimentally excluded, and are in general disagreement with spectroscopic data \cite{Serenelli:2016nms}. 

Finally the decay of $^8$B nuclei produced in the $pp$ and $pep$ chains yields the highest energy 
neutrinos, within the 1-10 MeV range. These are expected to produce nuclear recoils in DD
experiments near the $E_R \sim$ keV recoil energy threshold.   Even though the $^8 \rm{B}$
neutrino flux is six orders of magnitude lower than the $pp$ flux, the coherent enhancement of the cross section with the atomic number ($\sigma
\sim A^2$) significantly boosts the detection rate via nuclear 
scattering and implies that heavy target DD experiments may be sensitive to this signal.

\begin{figure}
\includegraphics[width=.5\textwidth]{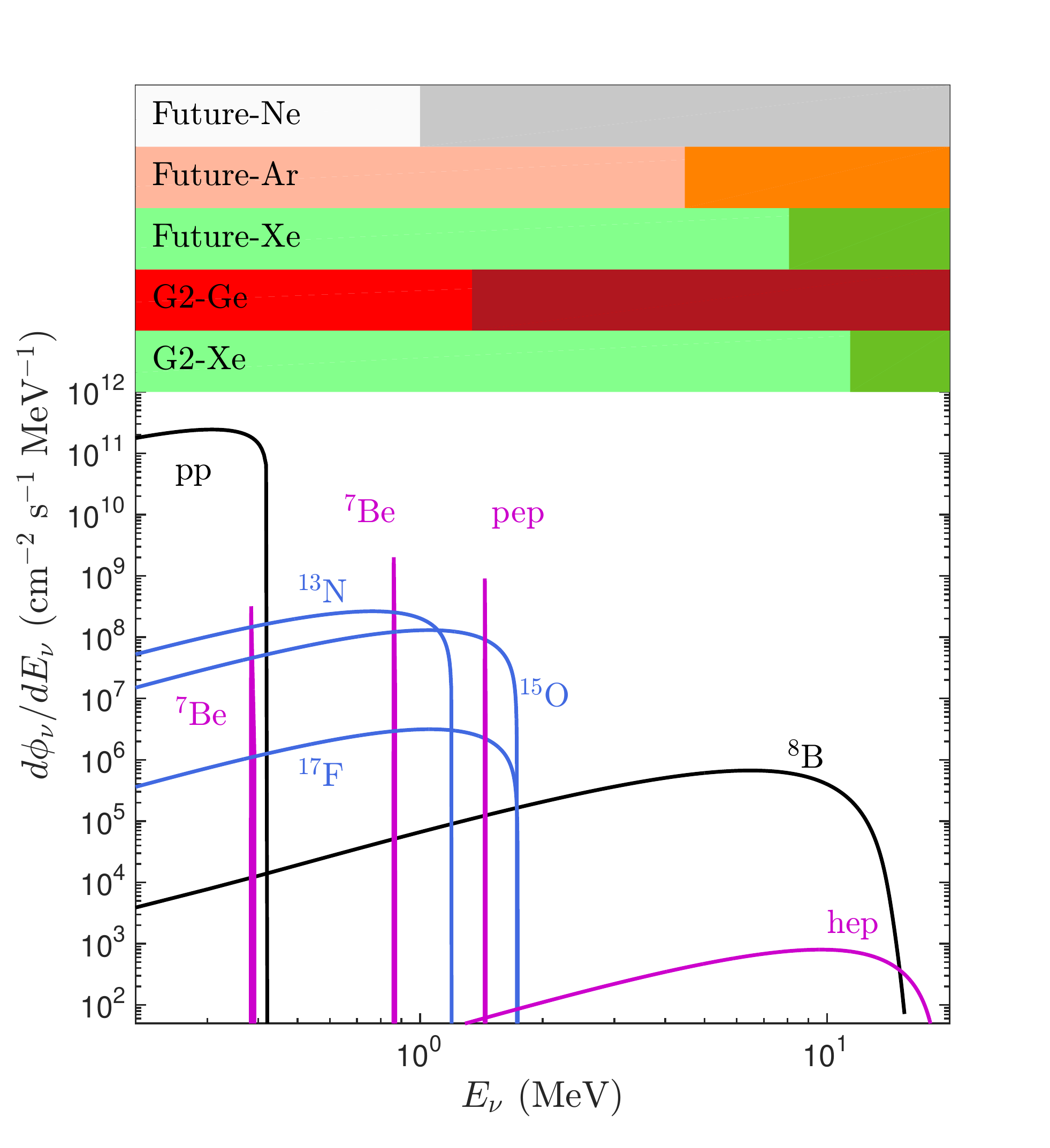}
\caption{Spectra of solar neutrinos accessible to direct detection experiments. In black are the $pp$ and $^{8}$B fluxes that will be seen respectively by electron and nuclear recoils in second generation (G2) and future experiments. CNO fluxes are in blue. The purple Be and pep lines, as well as the subdominant hep flux are not considered in this work. The bands at the top of the figure illustrate the reach of electron recoils (light shading) and nuclear recoils (dark shading) in future experiments, based on the optimistic configurations listed in Table \ref{Table:exps}. A low-threshold experiment with a light target nucleus may be able to probe the CNO fluxes for the first time, provided that backgrounds are low enough and nuclear recoils can be discriminated.}
\label{fig:neutrinofluxes}
\end{figure}

Fig.~\ref{fig:neutrinofluxes} shows the individual spectra for the solar neutrino fluxes mentioned above. We also use coloured bands to show the reach of the experiments that we consider in this work. Dark shading shows the neutrino energy range that can be seen via coherent nuclear scattering, while the light shaded areas show the reach of electron recoils. In reality, only the $pp$ spectrum is expected to lead to a visible  electron recoil signal: this is due to 1) the fact that $pp$ dominates the flux by four orders of magnitude at low energies; and 2) very high radioactive backgrounds expected at larger electron recoil energies, which will dwarf even the upper edge of the $pp$ spectrum (see e.g. Ref.~\cite{Baudis:2013qla}).

Note that atmospheric neutrinos and neutrinos from the diffuse supernova background 
could also induce nuclear recoil signatures in DD experiments. However, since they are produced at higher energies and with much lower rates, 
 they should only be within the reach of future multi-ton experiments. We disregard them in the present study.

\subsection{Neutrino physics} 

In the SM, the coherent scattering of solar 
neutrinos with nuclei (which takes place through the exchange of a $Z$
boson) typically leads to nuclear recoils below $10$~keV. Their detection thus requires  
 low-threshold detectors such as SuperCDMS.
On the other hand, neutrino-electron interactions (which occur through the exchange of neutral and
charged gauge bosons) give rise to electron recoil signatures 
of a few tens to hundreds of keV.

As the $pp$ flux was first measured by Borexino using electron recoils~\cite{borexinopp}, many authors have proposed the use of nuclear and electron recoil signatures in DD experiments to probe solar observables, test the validity of SM processes, and probe new physics at low
energies.

For example Ref.~\cite{Billard:2014yka} pointed out that a low-threshold
Ge detector could improve the measurement of the $^8$B flux
normalisation to better than 3\% and 
Ref.~\cite{Baudis:2013qla} showed that the observation of $pp$ neutrinos in a
 third generation Xe detector such as DARWIN \cite{Schumann:2015cpa} could
lower the statistical uncertainty on the $pp$ flux to less than 1\%; this is  
an overwhelming improvement over the current 10\% error from Borexino
\cite{borexinopp}. Such precision measurements can also help
distinguish between metal-rich and metal-poor solar models, via the
correlation between neutrino production and the environmental
abundance of primordial heavy elements \cite{PenaGaray:2008qe,
  Haxton:2009jh, 2013ARA&A..51...21H, Serenelli:2012zw}.

The nuclear recoil
event rates are sensitive to the weak (or Weinberg) angle $\theta_W$, which
expresses the ratio of the charged to neutral weak gauge boson masses,
\begin{equation}
  \cos \theta_W \equiv \frac{m_W}{m_Z},
\end{equation}
and effectively determines the ratio between the couplings of the
neutrino to the proton versus the neutron at low energies. The
quantity $\st$ has been determined to very high accuracy at the
electroweak scale, in high energy experiments.  Given LEP, PETRA and
PEP measurements~\cite{Agashe:2014kda,Behrend:1986md}, the SM
renormalization group equations imply that this parameter should run
to $\st = 0.2387$ at low energies in the $\overline{MS}$
scheme~\cite{Erler:2004in}. Thus far, the lowest-energy direct probe
of $\st$ has been at scales of $2.4$~MeV~\cite{Bouchiat:1983uf}, via
atomic parity violation measurements in $^{133}$Cs
\cite{Porsev:2009pr}. Given that the momentum exchange in coherent
neutrino-nucleus scattering occurs at energies of a few MeV, and that
electron recoils are expected to probe the $\mathcal{O}(10~{\rm keV})$
range, a direct measurement of $\st$ in future DD experiments would
constitute the first measurement of this quantity in the keV-MeV
range. 

Finally, precision measurements of solar neutrinos can help constrain
new physics contributions, including a sterile component in the solar
flux \cite{Billard:2014yka}, as well as the presence of new mediators,
particularly if they are light (below the GeV scale).  These light
mediators could have important consequences in neutrino
physics~\cite{Boehm:2004uq}, in the long standing proton radius
discrepancy~\cite{Batell:2011qq}, and in light DM
scenarios~\cite{Boehm:2002yz}.  Indeed, for sufficiently light
mediators, the scattering rate will grow as $1/q^2$ as one goes to
lower energies, so the low momentum transfer of DD experiments makes
them ideal laboratories for such searches.

\section{Neutrino scattering in DD experiments}
\label{sec:CNS}

Solar neutrinos might leave
a signal in DD experiments, either through their coherent scattering
with the target nuclei or through scattering with the atomic
electrons. 

\begin{table*}
\begin{tabular*}{\textwidth}{{@{\extracolsep{\fill} }c c c c c c c c}} 
\toprule
Experiment  & $\epsilon$ (ton-year) &  $E_{th,n}$ (keV) & $E_{th,o}$ (keV) & $E_{max}$ (keV) &  $R(pp)$ & $R(^8\mathrm{B})$ & $R(\mathrm{CNO})$ \\
\hline
G2-Ge & 0.25 & 0.35 & 0.05 & 50 & -- &[62 -- 85] & [0 -- 3] \\
G2-Si  & 0.025 & 0.35 & 0.05 & 50 & -- &[3 -- 3] & 0 \\
G2-Xe  & 25 & 3.0 & 2.0 & 30 & [2104 -- 2167] &[0 -- 64] & 0  \\
\hline
Future-Xe & 200 & 2.0 & 1.0  & 30 & [17339 -- 17846] &[520 -- 10094] & 0  \\
Future-Ar  & 150  & 2.0 & 1.0 & 30 &[14232 -- 14649] &[6638 -- 12354] &  0 \\
Future-Ne & 10 &  0.15 & 0.1 & 30 & [1141 -- 1143] &[898 -- 910] & [21 -- 63]\\
\hline
\hline
\end{tabular*}
\caption{Physical properties of idealized G2 (top 3 lines) and future
  experiments used in our forecasts, with the expected total $pp$ and
  boron-8 neutrino events, based on planned masses of similar experiments and an exposure of 5 years. We give
  nominal and optimistic threshold energies and maxima for the energy windows based on the energy beyond
  which backgrounds are expected to dominate. Our idealized
  G2 Ge and Si experiments are similar to the SuperCDMS
  SNOLAB phase, while the G2 Xe experiment is similar to LZ
  projections. Future experiments are similar to the planned DARWIN
  experiment, or an argon phase of a DARWIN-like experiment. 
}
\label{Table:exps}
\end{table*}

In general, the number of recoils per unit energy can be written
\begin{equation}
\frac{dR}{dE_R} =  \frac{\epsilon}{m_T} \int dE_\nu \frac{d\phi_{\nu }}{dE_\nu} \frac{d \sigma_{\nu }}{dE_R},
\end{equation}
where $\epsilon$ is the exposure and $m_{T}$ is the mass of the target electron or nucleus. If several isotopes are present, a weighted average must be performed over their respective abundances.

The SM neutrino-electron scattering cross section is
\begin{eqnarray}
\frac{d \sigma_{\nu e}}{dE_R} &=& \frac{G_F^2 m_e}{2 \pi}\biggl[ (g_v + g_a)^2  + \\
&& (g_v-g_a)^2\left(1-\frac{E_R}{E_\nu}\right)^2 + (g_a^2 - g_v^2)\frac{m_eE_R}{E_\nu^2} \biggr], \nonumber
\end{eqnarray}
where $G_F$ is the Fermi constant, and
\begin{equation}\label{eq:gv_ga}
g_{v;\mu,\tau} = 2\sin^2 \theta_W - \frac{1}{2}; \, \, \, g_{a; \mu,\tau}  = -\frac{1}{2},
\end{equation}
for muon and tau neutrinos. In the case $\nu_e + e \rightarrow \nu_e + e$, the interference between neutral and charged current interaction leads to a significant enhancement: 
\begin{equation}\label{eq:gv_ga_e}
g_{v;e} = 2\sin^2 \theta_W + \frac{1}{2}; \, \, \, g_{a;e}  = +\frac{1}{2}.
\end{equation}
The neutrino-nucleus cross section in the SM reads
\begin{equation}
\frac{d\sigma_{\nu N}}{dE_R} = \frac{G_F^2}{4 \pi} Q_v^{2} m_N \left(1-\frac{m_NE_R}{2E_\nu^2}\right) F^2(E_R) ,
\label{eq:nuesigma}
\end{equation}
where 
$F^2(E_R)$ is the nuclear form factor, for which we have taken the parametrisation given by Helm \cite{Helm:1956zz}.\footnote{Since we are mainly probing recoil energy regimes that are lower than typical DM searches, the uncertainty due to the choice of form factor is minimised \cite{Duda:2006uk}.}
$Q_v$ parametrises the coherent interaction with protons ($Z$) and neutrons ($N = A - Z$) in the nucleus:
\begin{equation}
Q_v  = N - (1-4 \st) Z .
\end{equation}

Current DD experiments excel at the discrimination of nuclear recoils from electron recoils. By design, these detectors
are engineered in such a way that the nuclear recoil background
induced by either radioactive processes or cosmic-rays is
extremely small. 
Thus, in our analysis we consider the idealised situation in which
nuclear recoils are produced solely by coherent neutrino scattering. This assumes that any nuclear recoil backgrounds can be completely identified and eliminated and that either no signal for dark matter has been found or that a potential dark matter background can be
discriminated.

On the other hand, electron recoils from radioactive processes are copious, and would constitute a very important background for the study of neutrino-electron scattering. Future advances in the design and construction of extremely radiopure detectors will allow a significant reduction of the noise levels. For example, current rates in Xenon100 electron recoil band are of the order of $3\times10^{3}\,\druty$ \cite{2011PhRvD..83h2001A}, but projected xenon-based experiments such as DARWIN aim to reduce this to ${\cal O}(10)\,\druty$ \cite{Baudis:2013qla} for recoil energies below 100 keV. 
In our analysis we will consider the idealized situation in which the electron recoil background is negligible compared to standard $\nu-e$ scattering.

For concreteness, we have specified in Table~\ref{Table:exps} several experiment types that are similar in threshold, efficiency and exposure specifications to upcoming experiments. We do not restrict ourselves to experiment-specific parameters such as background spectrum and resolution since these are difficult to estimate and subject to significant change. We thus include a second-generation germanium and silicon experiment (inspired by SuperCDMS SNOLAB), a second-generation xenon experiment (inspired by LZ), as well as future DARWIN-like xenon and argon experiments. Finally, we include a neon-based experiment to illustrate the possibility of observing the $^{15}$O and $^{13}$N neutrinos from the CNO cycle with future low-mass TPCs. The very recent Ref.~\cite{Strigari:2016ztv} contains some discussion of the $pep$ line; however, even for the most optimistic configuration that we consider, we would see at most 2 $pep$ events, versus a possible $\sim$ 60 CNO neutrinos in the same energy range.

Tab.~\ref{Table:exps} shows the parameters that we use for our benchmark models, and the expected number of events from electron recoils of $pp$ neutrinos, $R(pp)$, and nuclear recoils from $^8$B and CNO neutrinos ($R(^8B)$ and $R$(CNO), respectively). We have specified an exposure similar to planned experiments, as well as two sets of threshold energies that are respectively nominal and optimistic projections of what could be achieved in such experiments ($E_{th,n},
  E_{th,o}$). Last, as a stand-in for realistic efficiency curves, we take the efficiency in each experiment to rise linearly from 50\% at the threshold, to 100\% at 1~keV (for Ge, Si, Ne) or 5~keV (Xe, Ar). 

\section{Solar and Standard Model Physics}
\label{sec:solar}

The various components of the standard solar model (SSM) make use of very well-understood physics, but depend on over 20 individual input parameters. These include the solar age, luminosity, radial opacity dependence, diffusion rates, nuclear cross sections and the elemental abundances at age zero. 

Since the downward revision of photospheric elemental abundances a decade ago, some tension has remained between predictions of the SSM and independent observations using helioseismology. In this section, we focus on two parameters, the overall metallicity $Z/X$ and the effective change in opacity with respect to the SSM, $\delta\kappa$. With enough information, one should be able to study the effect of individual elements on the neutrino fluxes. However, with so few observables it is not possible to distinguish them.

We perform a Fisher analysis to extract the predicted sensitivity of future experiments to the various parameters studied here. For each experiment $k$ which measures an observable $\phi^k$ with error $\sigma_k$, the Fisher information matrix is
\begin{equation}
F^k_{ij} = \frac{1}{\sigma_k^2}\frac{\partial \phi^k}{\partial \theta_i}\frac{\partial \phi^k}{\partial \theta_j},
\end{equation}
where the indices $i,j$ run over the parameters $\{\theta\}$ that we wish to constrain. The total fisher matrix is simply $F \equiv \sum_k F^k$. Assuming gaussianity in the parameters of interest, the covariance matrix is
\begin{equation}
C = F^{-1}.
\end{equation}
The diagonal elements of $C$ are the forecasted errors on each individual parameter given the experiments included in $F$, while the off-diagonal components give the linear degeneracies.

For electron scattering described in Eq. \ref{eq:nuesigma}, the uncertainty on the neutrino mixing angles $\theta_{12}$ and $\theta_{13}$ lead to an extra source of uncertainty on the measured $pp$ flux. For G2 experiments, we take the $1\sigma$ errors on these parameters from the latest NuFit determinations\footnote{\url{http://www.nu-fit.org}} \cite{Gonzalez-Garcia:2014bfa,Bergstrom:2015rba}, which lead to a 1.15\% uncertainty on the inferred neutrino flux from neutrino scattering. We do not include this error for the Future experiments, as projects such as JUNO \cite{An:2015jdp} will constrain these quantities to very high precision\footnote{More concretely, JUNO expects to measure $\sin^2 \theta_{12}$ to within 0.67\%, leading to an error on the event rate in DD experiments of $\sim 0.2\%$. }$^,$\footnote{We also point out that by using independent measurements of the $pp$ flux, one can instead use the DD observations as a constraint on $\sin^2 \theta_{12}$. }.

\subsection{Neutrino fluxes and $\sin^2\theta_W$}

The lowest-energy sensitivity to $\st$ arises through neutrino-electron scattering, which probes interactions via momentum transfers of order tens of keV (though nuclear scattering recoil energies are lower, the transferred momentum $q = \sqrt{2 E_R m_N}$ is much higher). 

If only the experimental measurement by Borexino \cite{borexinopp} of the $pp$ flux is considered,
then we find that future DD experiments can measure $\st$ down to
about $20\%$ uncertainty. However, much greater precision can be
attained through the addition of the luminosity constraint on the
total neutrino flux from the Sun. Using the global bounds derived in
Ref. \cite{Bergstrom:2016cbh}, the resulting $0.6\%$ error on the $pp$
flux allows G2 experiments to narrow down the $\st$ measurement to
within 4.5\%. This is solely due to an LZ-like xenon experiment, as
the $pp$ flux will remain inaccessible to solid-state experiments due
to high backgrounds. Future liquid noble gas experiments can bring
this error down to 1.4\%. The projected uncertainties in different
configurations are given in Tab. \ref{tab:errors}. 

Lowering the
threshold has little impact on these numbers, since the electron
recoil rate is fairly insensitive to the lower energy. The expected
precision on the measurement of $\sin^2\theta_W$ is thus very close to
the results of present experiments, with the additional advantage that
direct detection experiments can access an energy range that is
unreachable in a collider setup, and is two orders of magnitude lower
than results from atomic parity violation experiments. As a final
  remark about $\st$, although the precision of future DD experiment
  would be about 10 times weaker than future experiments like
MOLLER~\cite{Mammei:2012ph}, the energy scale would be a factor
  10,000 smaller. Hence, DD experiments are sensitive to new physics at much lighter scales.

By instead fixing the value of $\st$ to the expected value given by running the LEP measurement down to low energies using the $\overline{\rm MS}$ scheme, $\st = 0.2387 \pm 7 \times 10^{-5}$
\cite{Agashe:2014kda}, the neutrino fluxes can be independently measured. One
can then predict the precision of the $^8$B and $pp$ flux
measurements from future experiments. The one-dimensional errors on each of these fluxes
are presented in the first column of Tab.~\ref{tab:errors}. The reduction of error in the pp flux is striking: the G2 xenon experiment will bring this from the current 10\% down to 2.2\%; future experiments bring this down even further, to 0.6\%\footnote{The small difference with the 1\% error quoted in Ref.~\cite{Baudis:2013qla} is due to the larger exposure we take here.}. Note that lowering the threshold has very little effect on the measured pp flux, as the electronic recoil rate does not rise sharply at lower energies. 

In contrast, a lower threshold allows significantly more $^8$B neutrinos to be measured, this time allowing a SuperCDMS-like germanium experiment to drive the G2 measurements, albeit with only a small improvement ($\pm 1.9$\%) with respect to current measurements ($\pm 2$\%). As a further consequence, the optimistic detector configurations have almost twice the sensitivity as the nominal ones. A xenon phase of DARWIN could thus measure the $^8$B flux better than even a dedicated future neutrino experiment such as HyperKamiokande \cite{Abe:2011ts}, for which we show a sensitivity projection based on one year of data taking in the last line of Tab.~\ref{tab:errors}. 

\begin{table}
\begin{tabular}{c | c c c}
Exp. & $\phiB$ & $\phipp$ & $\st$ \\ \hline
Measured & 2.0\% \footnote{Global fit \cite{Abe:2010hy}  $\phiB =(5.1 \pm 0.1) \times 10^6$ cm$^{-2}$ s$^{-1}$} & 10.6 \%\footnote{Borexino \cite{borexinopp} measurement  $\phipp = (6.6 \pm 0.7) \times 10^{10}$ cm$^{-2}$ s$^{-1}$}& \\ \hline
G2 			& 1.9\% (1.9\%)			& 2.5 \% 	(2.5\%)		& 4.6\% (4.5\%) \\
Future-Xe	& 1.8\% (0.9\%)		& 0.7\% (0.7\%) 				& 1.7\% (1.7\%) \\
Future-Ar 	& 1.0\% (0.6\%) 	& 0.6\% (0.5\%) 	& 1.5\% (1.4\%) \\ \hline
HyperK\footnote{Based on 1 year projected data \cite{Abe:2011ts}} & 1.43\% & -- & --\\\hline
\end{tabular}
\caption{Current and projected errors on the $^8$B and $pp$ neutrino fluxes (with fixed $\st=0.2387$), and on $\st$ at low energies (using the solar neutrino flux data from SuperK, SNO+ and Borexino, and the luminosity constraint on the $pp$ flux from \cite{Bergstrom:2016cbh}). The numbers are shown for the nominal (optimistic) thresholds of
  Tab.~\ref{Table:exps}. Each subsequent experiment set includes the
  previous one measurements.  \label{tab:errors}}
\end{table}

\subsection{Solar observables and CNO neutrinos}

\begin{figure*}
\includegraphics[width=.5\textwidth]{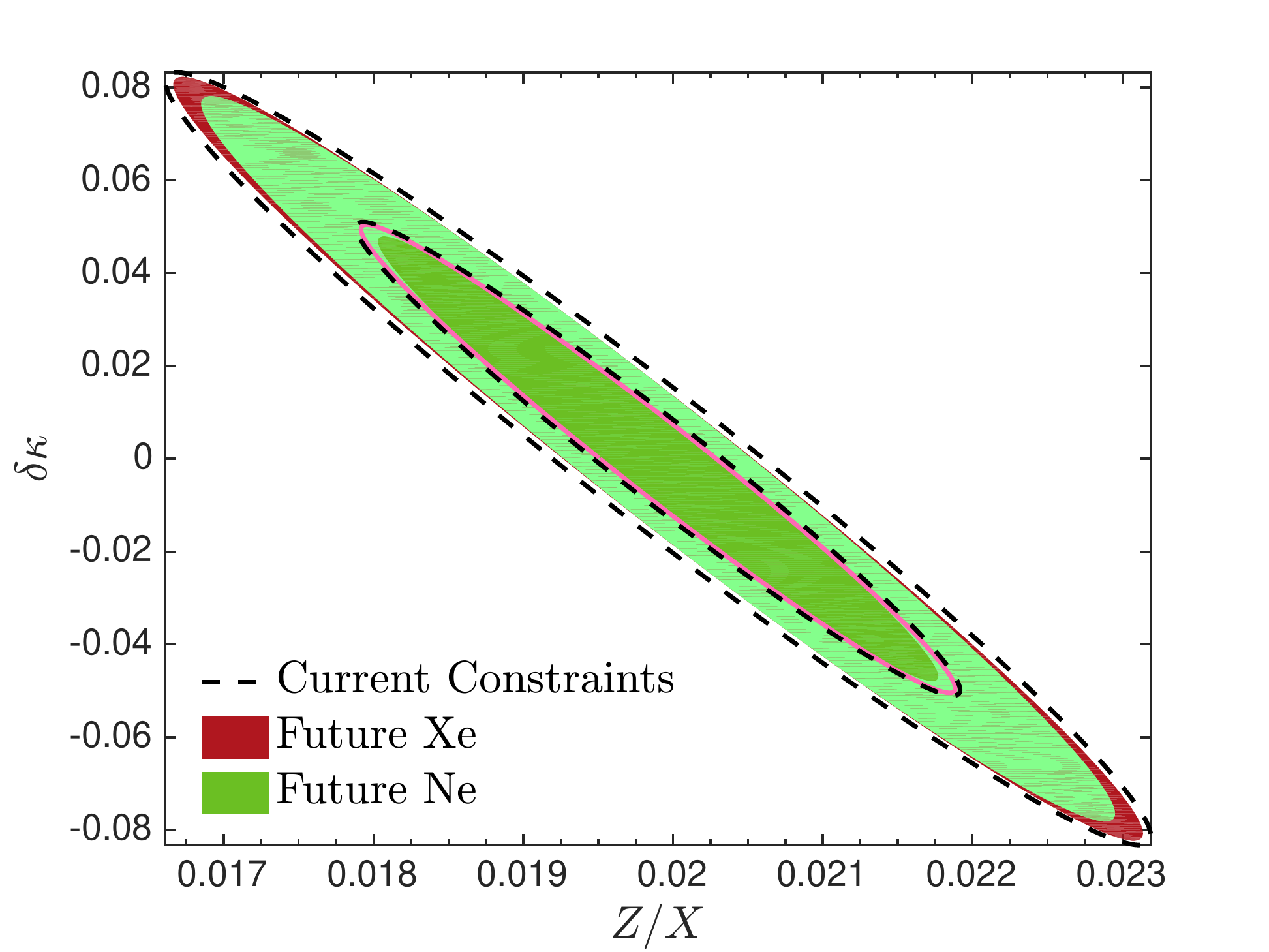}\includegraphics[width=.5\textwidth]{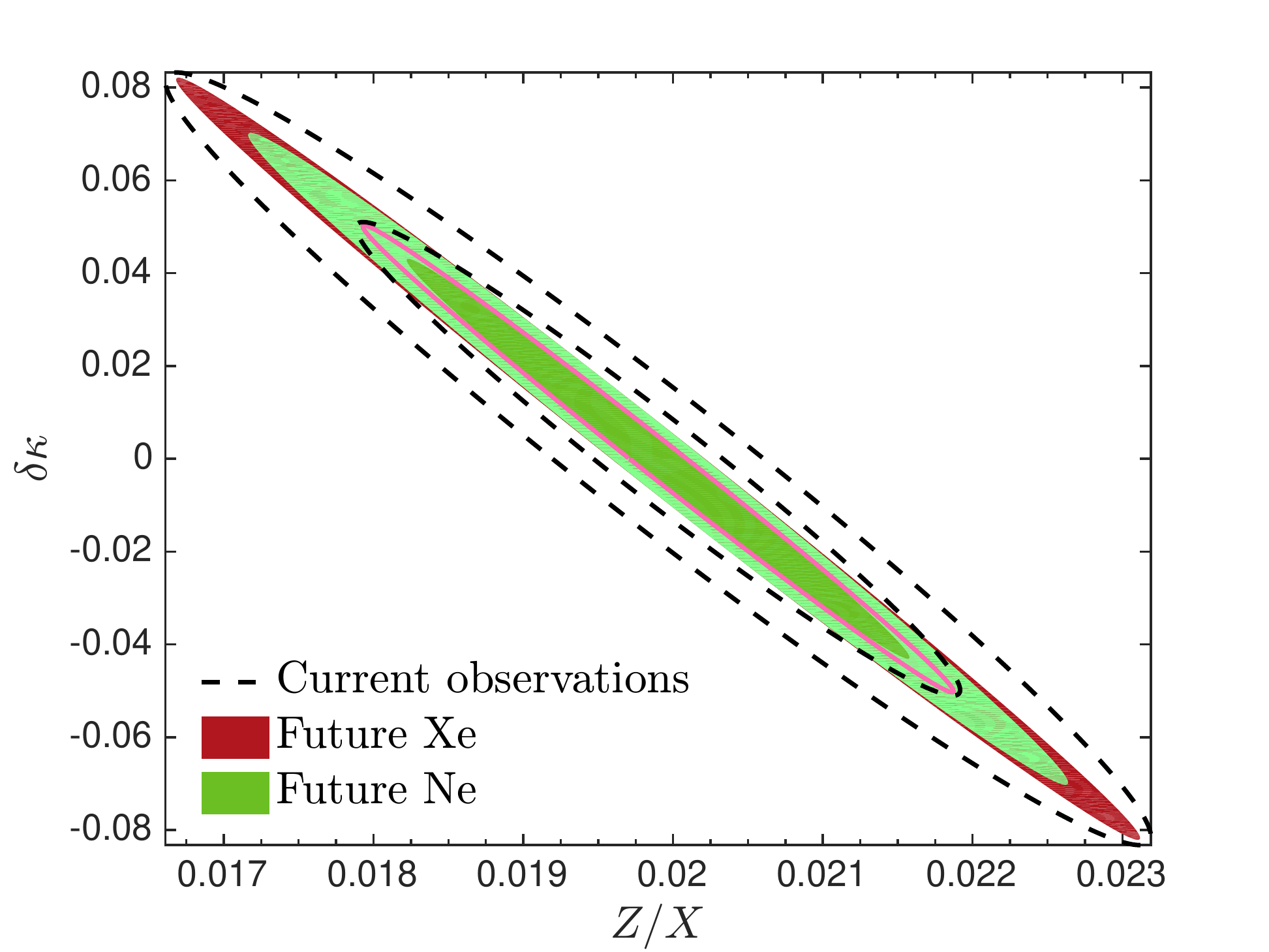}
\caption{Estimate of the potential improvement of constraints on the solar metallicity fraction $Z/X$ and the deviation of opacity from the standard solar model $\delta \kappa$, through the use of neutrino measurements.  Left: using the nominal thresholds $E_{th,n}$ from Table \ref{Table:exps}; right: using the optimistic threshold energies $E_{th,o}$. The main effect comes from the correlation between $\phiB$ and these two quantities (although the small effect of $pp$ is also included), based on linear solar models \cite{2010ApJ...724...98V,2005ApJ...626..530B}. The addition of these neutrino data are reflected in a narrower correlation line. To break this degeneracy, the neutrino flux from the subdominant CN cycle must be included, since these are the only direct probe of the light elements which contribute directly to $Z/X$. }
\label{fig:solarobs}
\end{figure*}

We now turn to the impact of coherent nuclear scattering and electron recoil measurements on solar modelling. An accurate measurement of $pp$ and $^7$Be neutrino fluxes, together with the luminosity constraint \cite{Bahcall:2001pf}, tell us the exact fraction of the solar energy that comes from the pp-chain, and thus indirectly constrains the CNO cycle which makes up the remaining $\sim 1\%$ of energy generation. Such a measurement would therefore constitute the most accurate probe yet of the $^{13}$N and $^{15}$O neutrino fluxes, which have yet to be experimentally observed.

\begin{table}
\begin{tabular}{c | c c }
Exp. & $\delta_k$ & $Z/X$  \\ \hline
Solar observations & 0.04 & 0.0024 \\
Including neutrinos & 0.034 & 0.0013  \\
\hline
Future-Xe	&  0.033 (0.033)	&	0.0013 (0.0013)			 \\
Future-Ne 	& 0.032 (0.030)	& 	0.0013 (0.0012) \\ \hline
\end{tabular}
\caption{Current and projected errors on the average opacity ($\delta_k$) and the metallicity ($Z/X$), using the nominal (optimistic) thresholds of Tab. \ref{Table:exps}. Each subsequent experiment set includes the previous one measurements. Approximate error on $\delta \kappa$ is from helioseismological observations  \cite{2010ApJ...724...98V}; error on the metallicity comes from the measured solar abundances  \cite{2009ARAandA..47..481A}. The neutrino data used for the second line are the same as Tab. \protect\ref{tab:errors}.\label{tab:errors_solar}}
\end{table}

To illustrate how DD experiments can help infer solar properties, we use the partial derivatives from linear solar models \cite{2005ApJ...626..530B,2010ApJ...724...98V} of the neutrino fluxes with respect to an overall shift in opacity $\delta \kappa$ and with respect to the metal-to-hydrogen ratio $Z/X$.\footnote{Strictly speaking, one should correlate all neutrino fluxes with individual elemental and with a radially-dependent parametrization of $\delta \kappa(r)$. Given the paucity of observables, however, we restrict ourselves to these more general quantities.} Fig.~\ref{fig:solarobs} illustrates the impact of future experiments on these observables. Since they are both highly correlated with the $^8$B flux (and less so to $pp$), the main effect is to narrow down the degeneracy in the opacity-metallicity space. This indicates that, when combined with new neutrino data, a precise independent measurement of either $\kappa$ or solar metallicity can pin down the other observable to a high degree of precision, though direct detection experiments on their own are of limited help. In Tab.~\ref{tab:errors_solar}, we show the projected errors from DD experiments on the determination of opacity and metallicity.

We finally return to the $^{13}$N and $^{15}$O neutrino fluxes. These lie approximately an order of magnitude below the $^7$Be line, although they dominate the solar spectrum at electron recoil energies above $\sim 750$ keV. In this range, material backgrounds dominate by several orders of magnitude, making discrimination of the CNO neutrinos highly unlikely. One may search for CNO events in the nuclear recoil spectrum. This requires:
\begin{enumerate}
\item A light target to reach neutrino energies below the $^8$B peak;
\item A low threshold (e.g., below 0.25 keV for Ne);
\item Good discrimination between nuclear and electron recoil events.
\end{enumerate}
Light elements such as F, Na, Ne and Al, would give 0.02 to 0.01
$^{15}$O events per kg year with a 0.1 keV threshold. Superheated
fluids such as C$_3$F$_8$ targets are light enough, but current thresholds
tend to be far too large \cite{Amole:2016pye}. Supercooled scintillating sapphire
bolometers such as ROSEBUD \cite{Bobin:1999yy} can attain a $\sim 100$ eV
threshold. However, their exposures tend to be limited to a few
hundred gram-days. An alternative which satisfies 1) and 2) would be
pressurized noble gas TPCs. Thresholds of 100 eV can in principle be
obtained for Ne with such a setup \cite{Iguaz:2015myh,Irastorza:2015geo}. The low threshold is key to this observation:
increasing $E_{th}$ by only 50\% yields a substantial drop in the
observable CN flux, from 63 to 21 events for 10 ton-years of
neon. Discrimination
  between nuclear and electron recoils at such low energies is
  extremely challenging and one might rather rely on the radiopurity
  of the experiment and the directionality of solar neutrinos. In
green, we show the effect of a future neon-based experiment on solar
observables in Fig. \ref{fig:solarobs}. CNO neutrinos are a direct probe of the abundances of
these elements -- and thus of the metallicity $Z/X$ which C, N, O
dominate. Even a modest measurement of the CNO flux (63 events in our
optimistic case) can thus provide a small break in the degenerate
parameter space.

\section{Constraints on new physics}
\label{sec:newphysics}

The simultaneous
measurement of the neutrino-electron and neutrino-nucleus couplings
enables us to test the predictions for models of new physics in two
complementary directions.

\subsection{Simplified models}

\begin{table*} [!t]
\begin{tabularx}{\textwidth}{l | X | X | X} 
\toprule
Mediator \rule{0ex}{3ex}  \raisebox{-1ex}{\rule{0ex}{3ex}}  
& $ \mathcal{L}$ & $d\sigma_e/dE_R-d\sigma^{\rm SM}_e/dE_R$ 
& $d\sigma_N/dE_R-d\sigma^{\rm SM}_N/dE_R $\\
\hline
Scalar \rule{0ex}{4ex}  \raisebox{-5ex}{\rule{0ex}{3ex}} 
& $\left(g_{\nu,\phi}\,\phi \bar{\nu}_R \nu_L + h.c.\right) \newline + \phi \bar{\ell} g_{\ell,s}  \ell +  \phi \bar{q}  g_{q,s}  q $
&$\frac{g_{\nu ,\phi }^2 g_{e,s}^2 E_R m_e^2 }
      {4 \pi E_{\nu }^2 \left(2 E_R m_e+m_\phi^2\right)^2}$
& $\frac{Q_s^{\prime 2} m_N^2 E_R  }{4 \pi E_{\nu }^2 \left(2 E_R m_N+m_\phi^2\right)^2}$ \\ \hline
Pseudoscalar \rule{0ex}{4ex}  \raisebox{-5ex}{\rule{0ex}{3ex}} 
&  $\left(g_{\nu,\phi}\,\phi \bar{\nu}_R \nu_L + h.c.\right) \newline -i\gamma^5 \phi \bar{\ell} g_{\ell,p}  \ell -i\gamma^5  \phi \bar{q}  g_{q,p}  q $
&$\frac{g_{\nu ,\phi }^2g_{e,p}^2 E_R^2 m_e}
      {8 \pi E_{\nu }^2 \left(2 E_R m_e+m_\phi^2\right)^2}$
& 0\\ \hline
Vector \rule{0ex}{4ex}  \raisebox{-10ex}{\rule{0ex}{3ex}} 
& $g_{\nu,Z'} Z'_\mu \bar{\nu}_L \gamma^\mu \nu_L \newline + Z'_\mu \bar{\ell} \gamma^\mu  g_{\ell,v} \ell +Z'_\mu \bar{q} \gamma^\mu  g_{q,v}  q$ 
 &$ \frac{\sqrt{2}G_Fm_e g_vg_{\nu,Z'}g_{e,v}}{\pi\left(2 E_R m_e+m_{Z'}^2\right)} \newline
  +\frac{m_e g_{\nu,Z'}^2g_{e,v}^2}{2\pi\left(2 E_R m_e+m_{Z'}^2\right)^2}$ 
&$-\frac{G_F m_N Q_v Q'_v (2E_\nu^2-E_Rm_N)}{2\sqrt{2}\pi E_\nu^2\left(2 E_R m_N+m_{Z'}^2\right)} 
\newline +\frac{Q_v^{\prime 2}m_N(2E_\nu^2-E_Rm_N)}{4\pi E_\nu^2\left(2 E_R m_N+m_{Z'}^2\right)^2} $ \\ \hline
Axial Vector \rule{0ex}{4ex}  \raisebox{-10ex}{\rule{0ex}{3ex}} 
& $ g_{\nu,Z'} Z'_\mu \bar{\nu}_L \gamma^\mu \nu_L \newline - Z'_\mu \bar{\ell} \gamma^\mu g_{\ell,a}\gamma^5 \ell \newline -Z'_\mu \bar{q} \gamma^\mu g_{q,a}\gamma^5 q$ 
& $\frac{\sqrt{2}G_Fm_e g_a g_{e,a}g_{\nu,Z'}}{\pi\left(2 E_R m_e+m_{Z'}^2\right)}\newline
  +\frac{m_e g_{\nu,Z'}^2g_{e,a}^2}{2\pi\left(2 E_R m_e+m_{Z'}^2\right)^2}$ 
& $\frac{ G_F m_N Q_a Q'_a(2E_\nu^2+m_N E_R)}{2\sqrt{2}\pi E_\nu^2\left(2 E_R m_N+m_{Z'}^2\right)} 
\newline - \frac{ G_F m_N Q_v Q'_a E_\nu E_R}{\sqrt{2}\pi E_\nu^2\left(2 E_R m_N+m_{Z'}^2\right)}
\newline +\frac{Q_a^{\prime 2} m_N (2E_\nu^2+E_Rm_N)}{4\pi E_\nu^2\left(2 E_R m_N+m_{Z'}^2\right)^2} $\\
\hline
\hline
\end{tabularx}
\caption{New Lagrangian terms and differential cross sections with the
  nucleus $N$ and electron $e$ for the four types of new mediator we
  consider. Note the negative interference in the vector and axial
  case with the SM contribution. The couplings $g_v$ and $g_a$ are
  defined in Eq.~(\ref{eq:gv_ga}). The coherence factors $Q_i$ are
  defined in Eqs.~(\ref{eq:coh-facs1}-\ref{eq:coh-facs2}).}
\label{Table:newphys}
\end{table*} 

To parametrise new physics at a very low scale, we write the
Lagrangian for energies below electroweak symmetry breaking which
incorporates scalar and vector mediators that couple to neutrinos,
electrons, and quarks in a model independent way. The additional terms
in the low energy Lagrangian are shown in the second column of Table
\ref{Table:newphys}, for a scalar (or pseudoscalar) mediator $\phi$,
and vector (or axial vector) mediator $Z'$.  The resulting
neutrino-nucleus and neutrino-electron scattering cross sections are
shown in the third and fourth columns. We neglect terms of order
$E_R/E_\nu \lesssim 10^{-2}$ and, in the case of pseudoscalar interactions, couplings to heavy quarks.

In our model, we introduce couplings to quarks, but in order to
correctly describe scattering with nuclei, the couplings of the
mediator to the nucleons are needed. This is done by calculating the
matrix element of the quark Lagrangian with nucleon states, which
leads to the following changes at the Lagrangian level:
\begin{eqnarray}
c_q \bar{q}   q   & \to & f_{Tq}^{(\mathcal{N})} \frac{m_{\mathcal{N}}}{m_q} \bar{\mathcal{N}} \mathcal{N}, \\
c_q \bar{q} i\gamma^5  q   & \to & c_A^{\mathcal{N}} \bar{\mathcal{N}} i\gamma^5 \mathcal{N},\\
c_q\bar{q} \gamma^\mu \gamma^5 q  & \to &   \Delta_q^{(\mathcal{N})} \bar{\mathcal{N}} \gamma^\mu \gamma^5 \mathcal{N}, \\
c_q \bar{q} \gamma^\mu q  & \to & c_V^\mathcal{N} \bar{\mathcal{N}} \gamma^\mu \mathcal{N},
\end{eqnarray}
where $\mathcal{N}$ is the nucleon (proton or neutron) spinor, and the coefficients are given numerically in 
Ref.~\cite{DelNobile:2013sia} (see also Ref.~\cite{Hill:2014yxa} 
and references therein, as well as Refs.~\cite{Alarcon:2011zs,Alarcon:2012nr} for more recent determinations of the pion-sigma term and of the strangeness content of the nucleon based on experimental and EFT results, which enter into these coefficients). More specifically, in our framework where 
the couplings to all quarks are the same, the coherence factors, $Q$, of 
the cross sections induced by these different interactions 
(see Table~\ref{Table:newphys}) are 
\begin{align}\label{eq:coh-facs1}
  &\frac{Q^\prime_s}{g_{\nu,\phi}g_{q,s}}=\sum_{\mathcal{N},q}\frac{m_\mathcal{N}}{m_q}f_{Tq}^{(\mathcal{N})}\approx 14A+1.1Z,\\
  &\frac{Q^\prime_v}{g_{\nu,Z'}g_{q,v}}=3A,\\
  &\frac{Q^\prime_a}{g_{\nu,Z'}g_{q,a}}=S_N\sum_q\Delta_q^{(p)}\approx 0.3 S_N,\\
  &Q_v=N-(1-4\st)Z,\\
  &Q_a=S_N(\Delta_u^{(p)}-\Delta_d^{(p)}-\Delta_s^{(p)})\approx 1.3 S_N. \label{eq:coh-facs2}
\end{align}
The primed coherence factors refer to new interactions. 
Although the axial vector interaction at low energies is also
coherent, it couples
to the spin operator, so the coherence factor is 
proportional to the nuclear spin $S^2_N \sim \mathcal{O}(1-10)$, rather than $A^2 \sim 10^4$.
This assumes a simple shell model, whereby any
unpaired nucleon contains the full $J$ quantum number of the nucleus
in its ground state.

We assume that the mediators are light (below the few GeV scale) and
their couplings to SM particles are small. Therefore, their
contribution to electron and nucleus scattering (via t-channel
exchange) should be negligible at a high momentum transfer $q^2\gg
m_{\phi,Z'}^2$ but will be enhanced for low scale measurements.

\subsection{Predicted event rates and sensitivities}

\begin{figure*}
\begin{tabular}{c c}
\includegraphics[width=.45\textwidth]{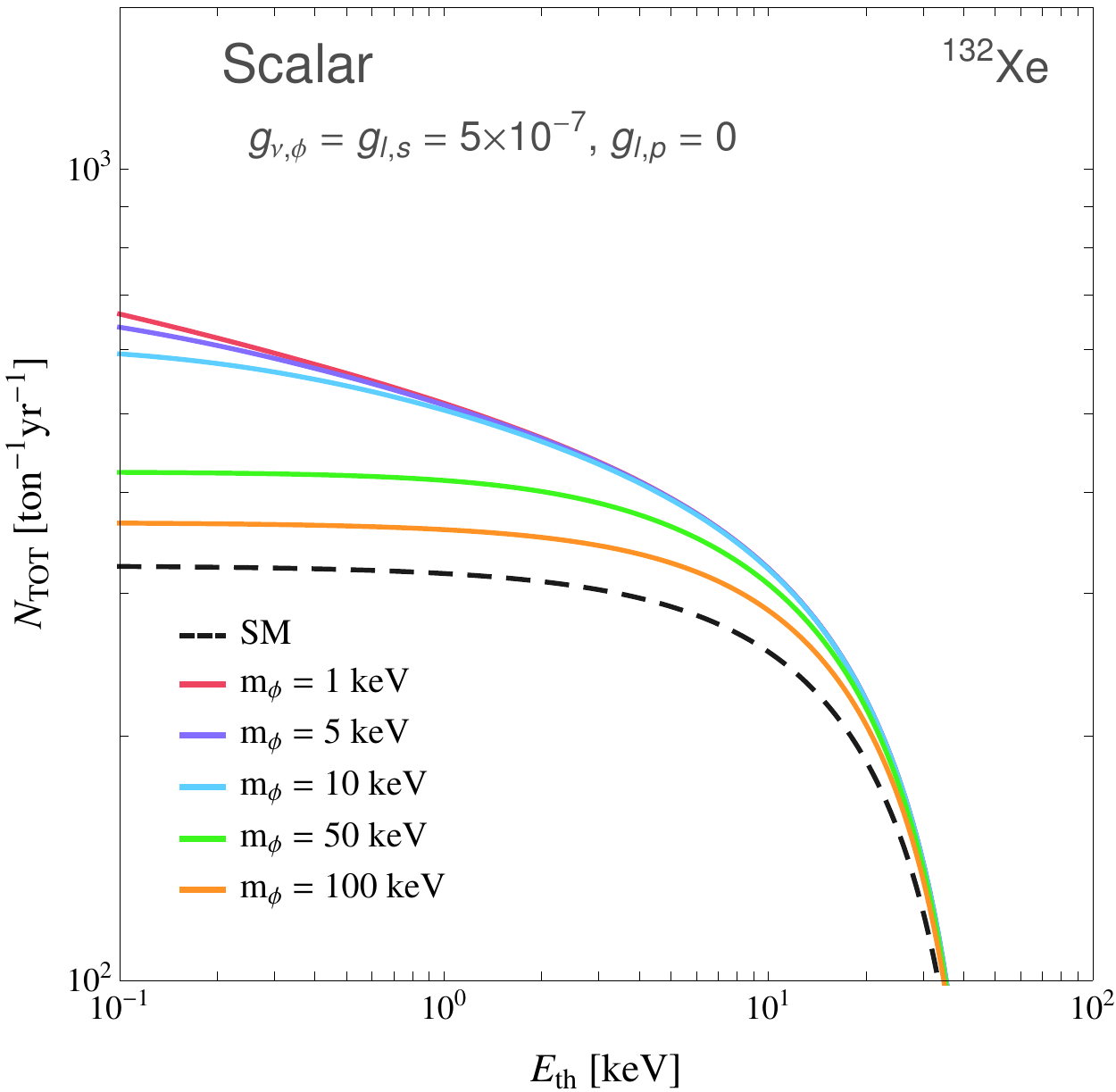} & \includegraphics[width=.45\textwidth]{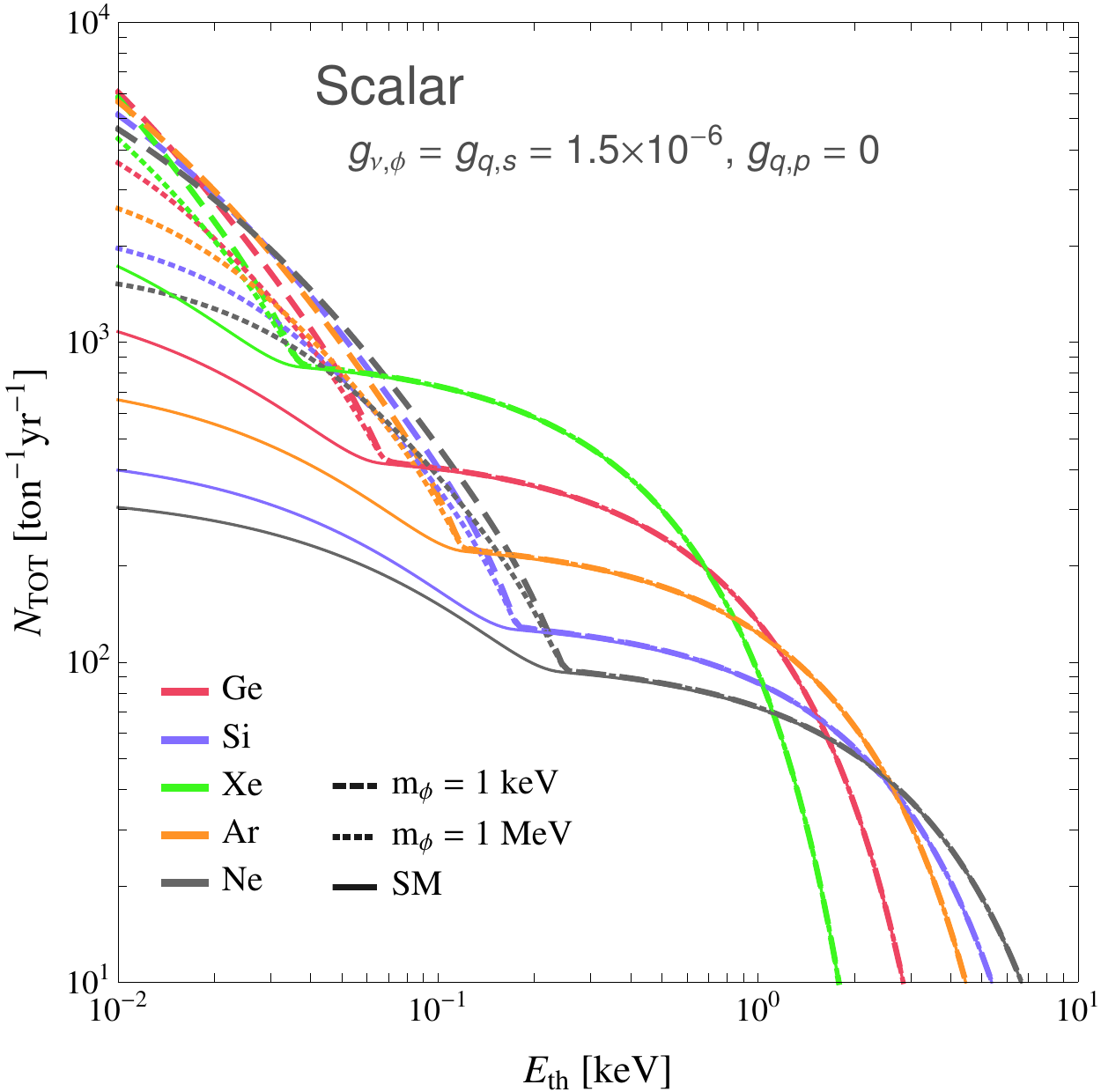} \\ 
\includegraphics[width=.45\textwidth]{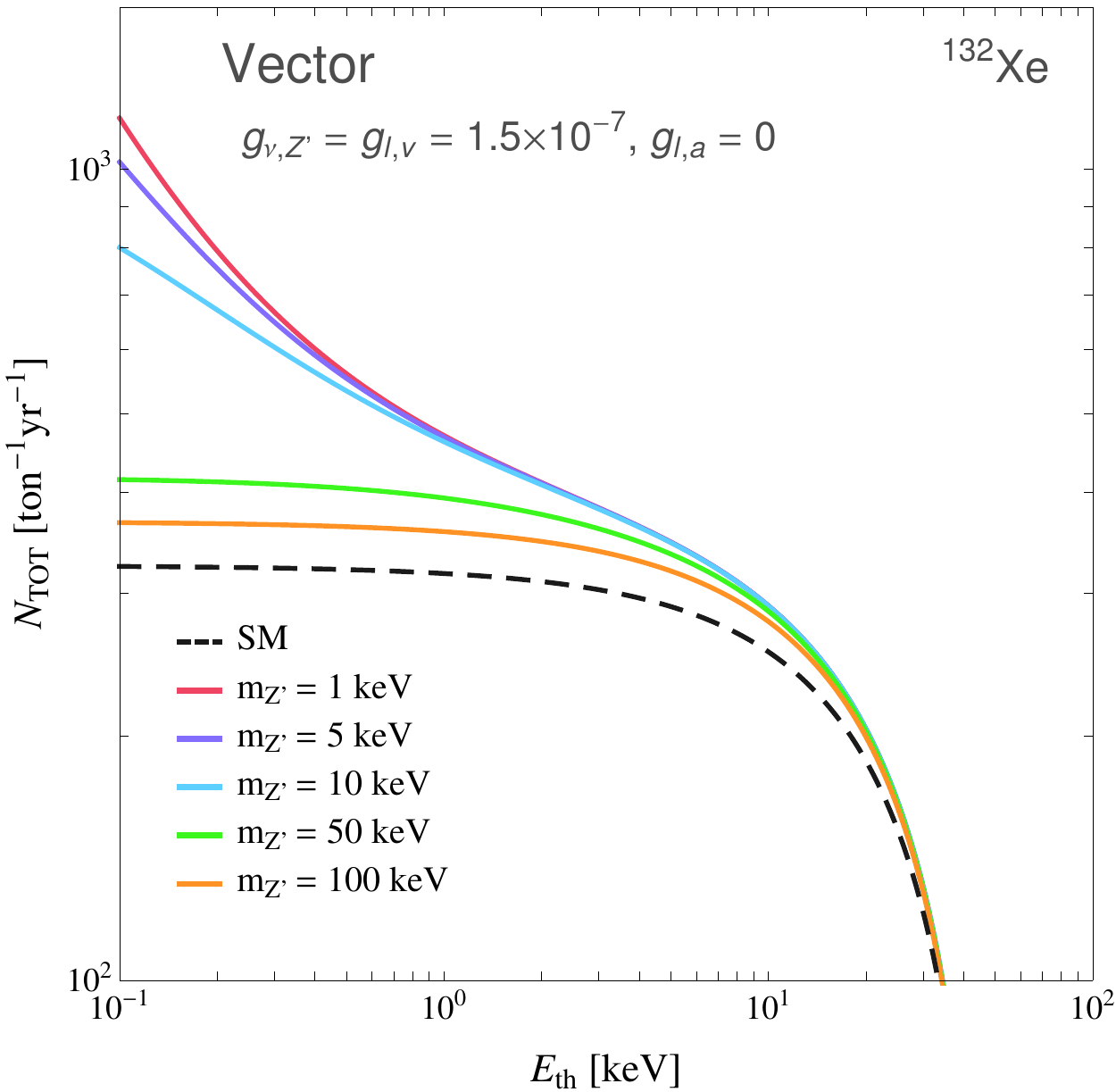} & \includegraphics[width=.45\textwidth]{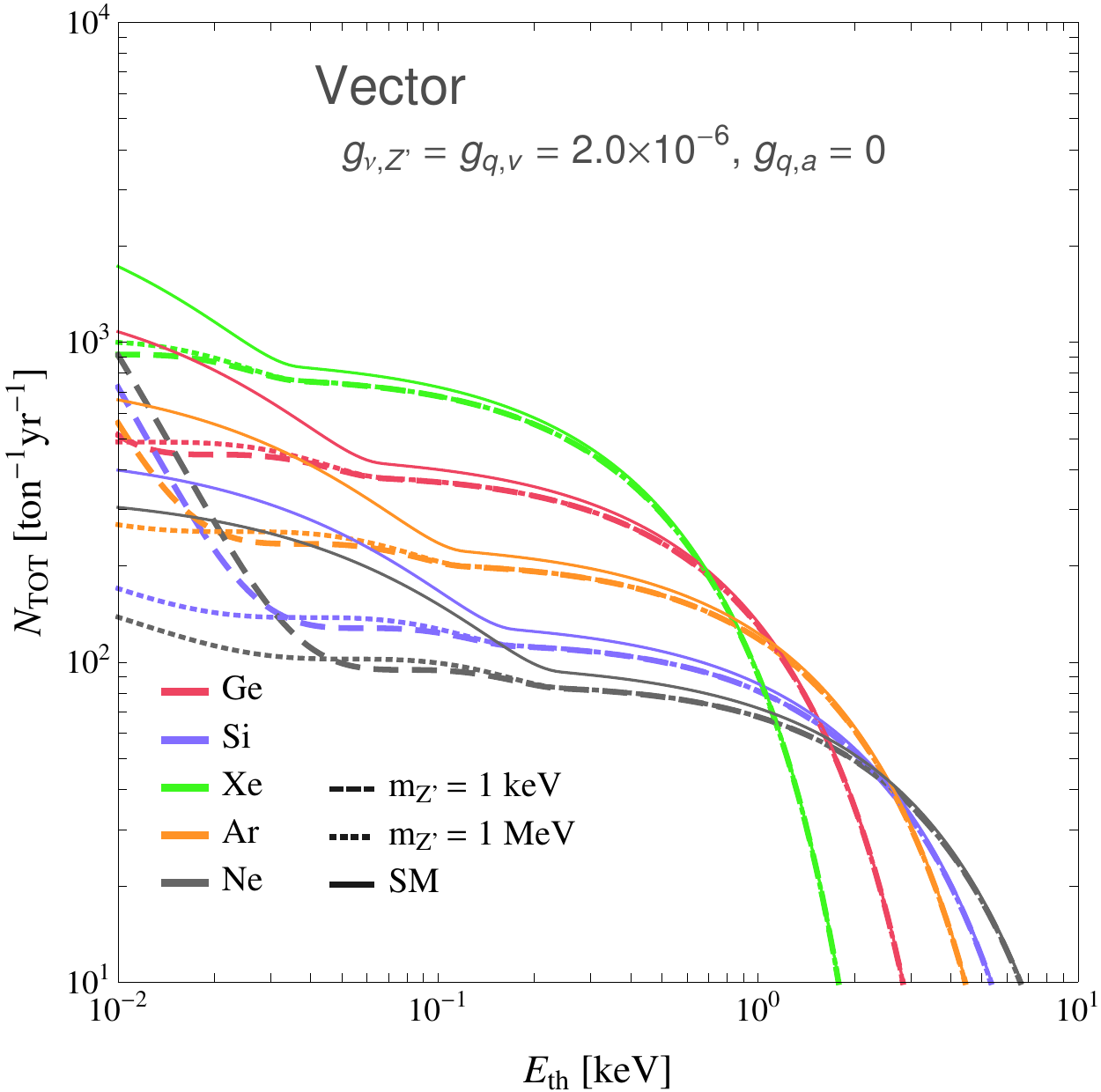}  \\
\includegraphics[width=.45\textwidth]{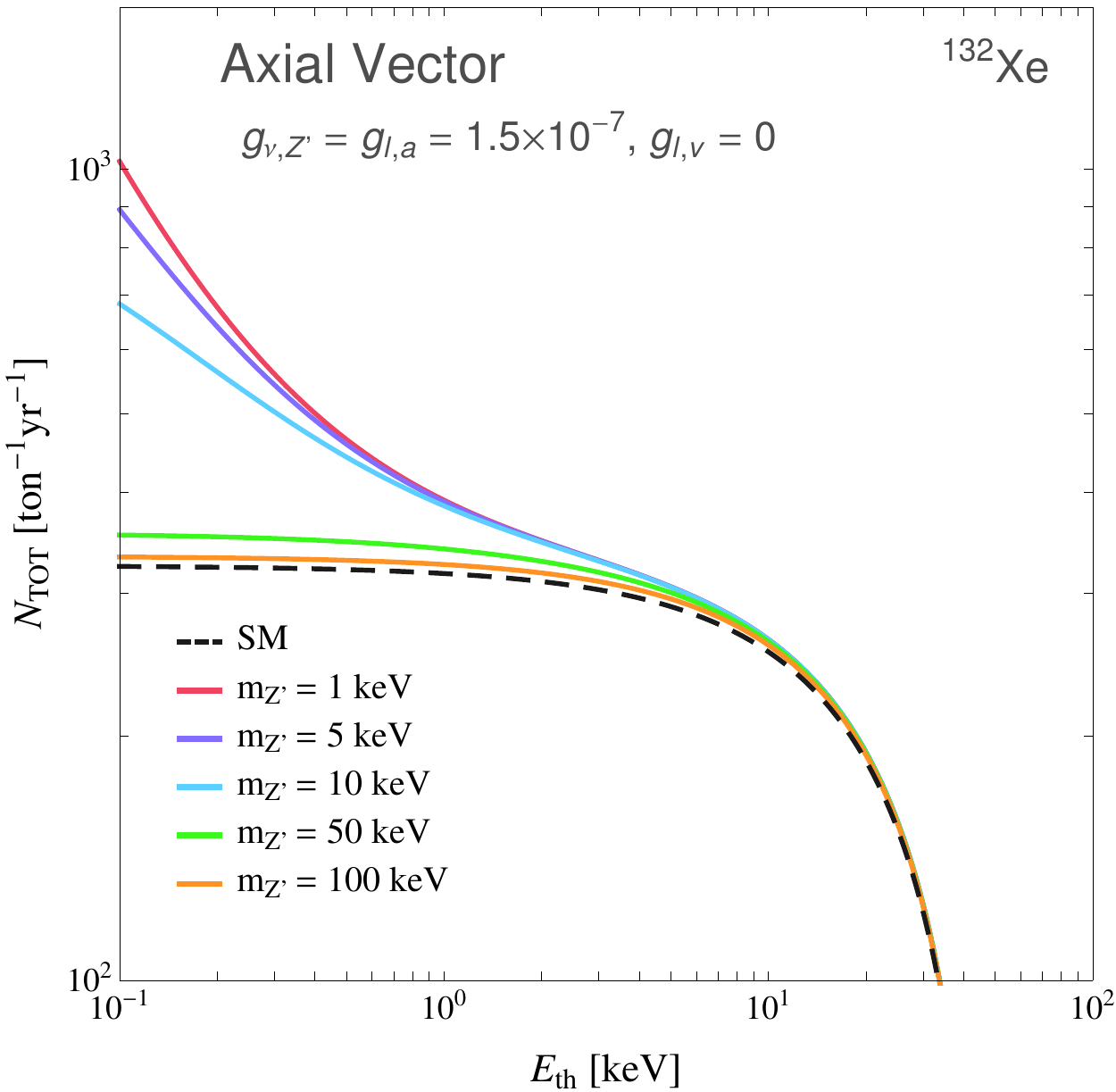} & \includegraphics[width=.45\textwidth]{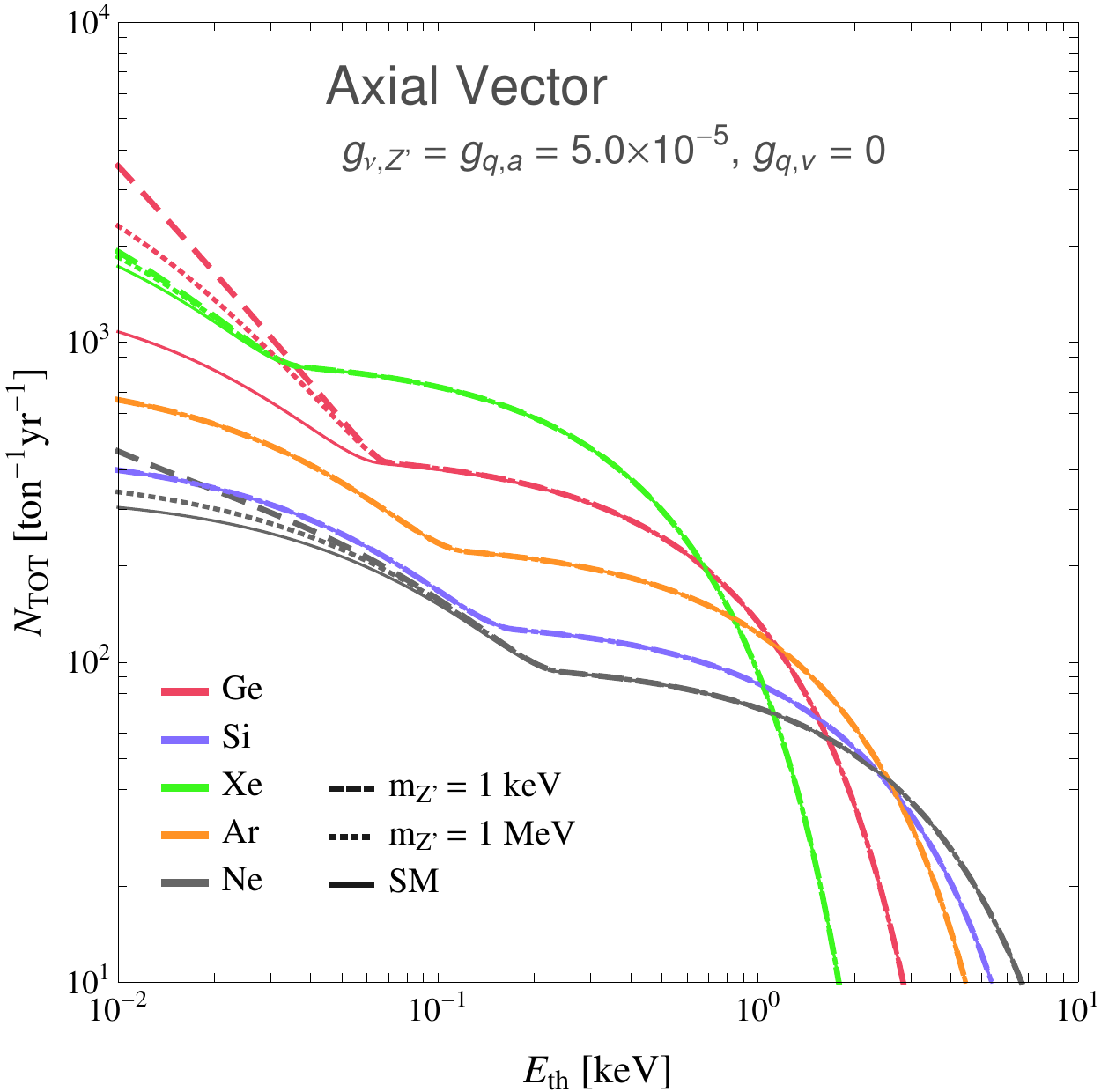}  
\end{tabular}
\caption{Electron recoil (left) and nuclear recoil (right) integrated rates as a function of the experimental threshold energy $E_{th}$.
  Electron recoils are normalised to $^{132}$Xe while nuclear recoils
  are plotted for a variety of target materials. Top: scalar coupling;
  middle row: vector coupling; lower panels: axial vector
  coupling.\label{newbosthresh}}
\end{figure*}

In Fig.~\ref{newbosthresh} we show the effect that the presence of scalar,
vector and axial vector interactions would have upon the rate of
scattering events per ton-year as a function of the low-energy
threshold.  The rate of electron
recoil events for a $^{132}$Xe target, as well as coherent nuclear recoil events for a
variety of different target materials and mediator masses are plotted. 
In all cases shown, the new
physics contribution grows with lower recoil energies, 
showing the need for low-threshold detectors.

Electron recoil spectra (shown on the left column) are from $pp$ neutrinos, the lowest energy and most copious neutrinos produced in the Sun. Since lowering the
threshold of detection does not open up any new sources of neutrinos,
a threshold of $E_{th}\sim 1$~keV is sufficient to maximize the SM
event rate.
The size of the new-physics contribution is dictated by the mass of the mediator and the corresponding coupling. In the limit of small mediator masses, the differential cross section in Tab.~\ref{Table:newphys} scales as $d\sigma/d E_R\propto E_R^{-1}$ for scalar mediators and $d\sigma/dE_R\propto E_R^{-2}$ for vector and axial vector mediators, thus leading to substantial changes at low energies. Therefore, if the experimental threshold is low enough, an enhancement of the signal with respect to the SM prediction could be observed. 
This does not hold for pseudoscalar mediators, as shown in Fig.~\ref{fig:pseudoscalar}, since in the small mass limit $d\sigma/d E_R$ is energy-independent. 
Although we are only showing the results for $^{132}$Xe, the rates for any other target can be found by rescaling by the corresponding number of electrons per unit mass.

For nuclear recoils the integrated event rate also increases
sharply with decreasing threshold. This can be seen as a sharp break in the right-hand panels of Fig.~\ref{newbosthresh}. This break corresponds to the intersection of new physics and SM contributions and its location depends on the values of the couplings. The fact that this enhancement becomes visible in these figures around the same energy as the CNO flux is a coincidence due to the choice of coupling, but the CNO contribution nonetheless results in further enhancement. 

The target material dependence is
very pronounced due to kinematics, as the maximum recoil energy is
suppressed by the large nucleus mass,
\begin{equation}
E_{R,\mathrm{max}}=\frac{2E_\nu^2}{(m_N+2E_\nu)}\ ,
\end{equation} 
For this reason, heavier targets need a lower
threshold to probe both new fluxes of neutrinos and new
physics processes at low energies. For example, whereas
$E_{th}\approx2$~keV is needed for xenon to be sensitive to $^8$B
neutrinos, these can be accessed by a hypothetical detector based on
neon with only $E_{th}\approx10$~keV.  The material dependence also
enters into the coherence factors (Eqs. \ref{eq:coh-facs1}--\ref{eq:coh-facs2}) for nuclear recoils which in turn depend on $A/Z$.

As in the case of electron recoils, the most pronounced deviations from the SM prediction occur in the limit where the mediator mass is small. Indeed, in such a case, the differential cross section scales as $d\sigma/dE_R\propto  E_R^{-1}$ for scalar mediators
and $d\sigma/dE_R\propto E_R^{-2}$ for vector and axial vector mediators when the new physics contributions dominate.
Once more, this leads to an enhancement of the cross section for low recoil energies. We do not show the nuclear recoil rates expected for a pseudoscalar mediator, since the nuclear form factor cancels out when the couplings to all light quarks are identical~\cite{DelNobile:2013sia}.

As an additional remark on the axial vector and vector mediator cases, the interference between the standard $Z$ and $Z'$ amplitudes become important when these are comparable in magnitude. Remarkably, this interference is destructive due to the chiral structure of the $Z$ couplings, which may lead to an overall suppression of events with respect to the SM prediction. We have illustrated this possibility in Fig.~\ref{newbosthresh} for the case of vector couplings.

\begin{figure*}
\begin{tabular}{c c}
\includegraphics[width=.45\textwidth]{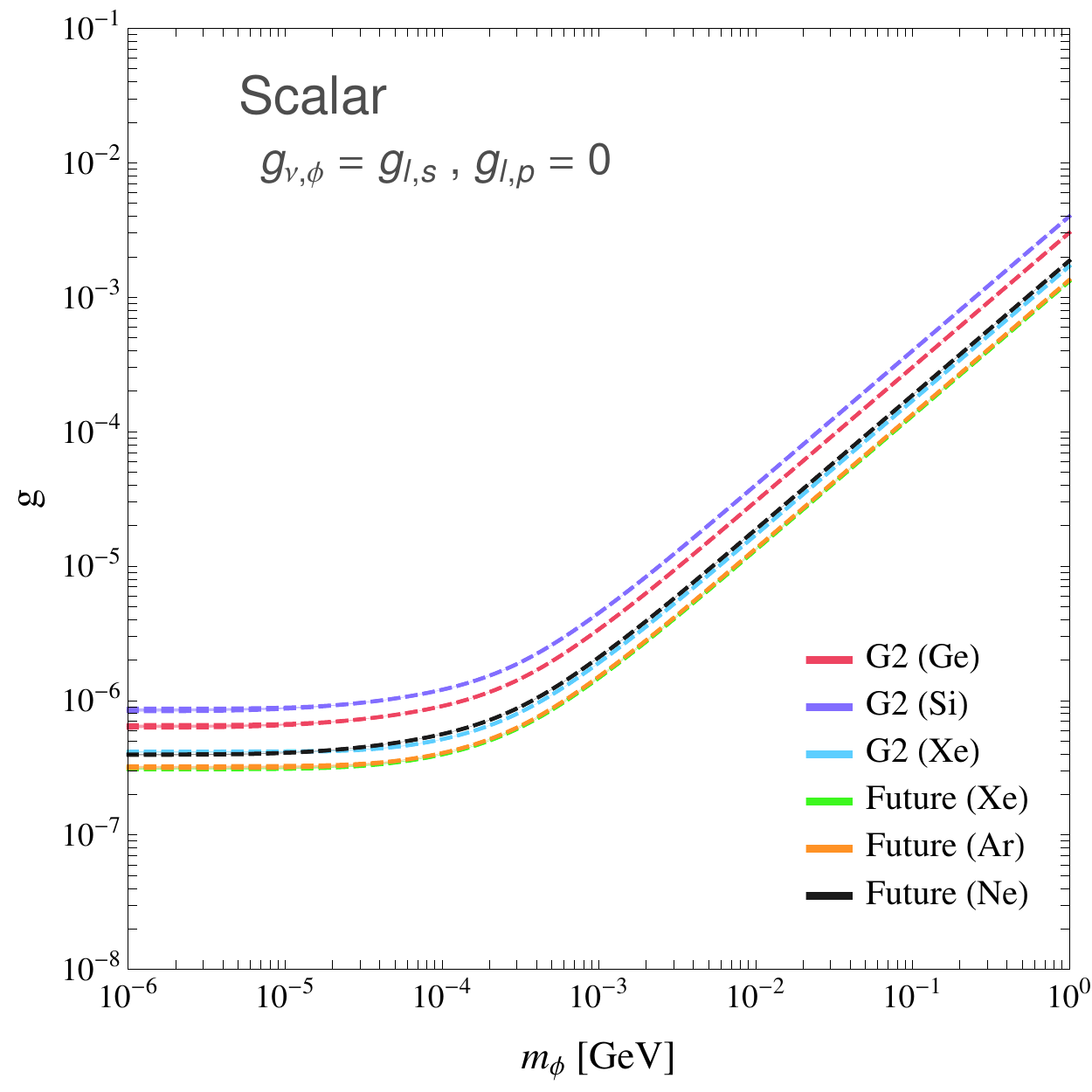} & \includegraphics[width=.45\textwidth]{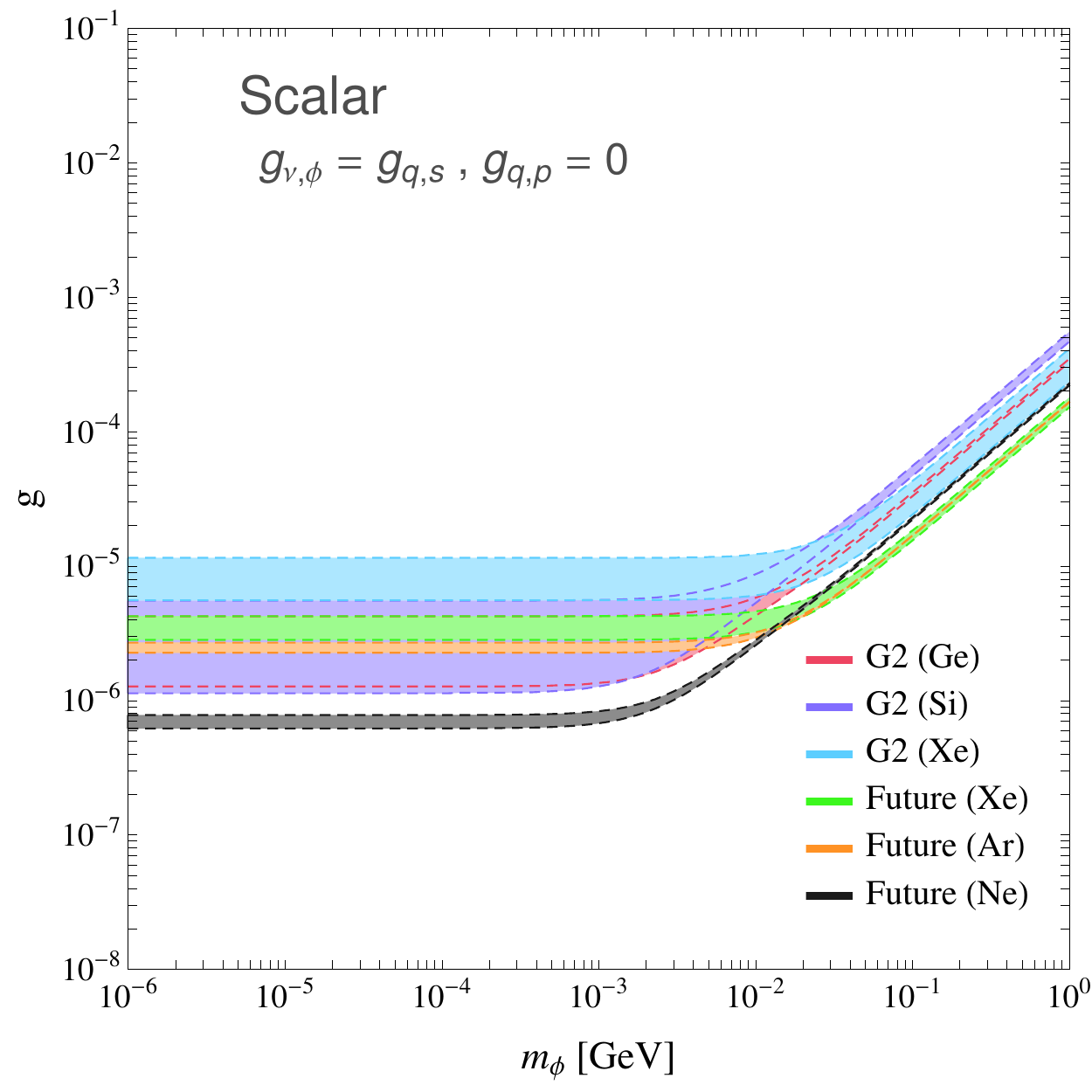}\\ 
\includegraphics[width=.45\textwidth]{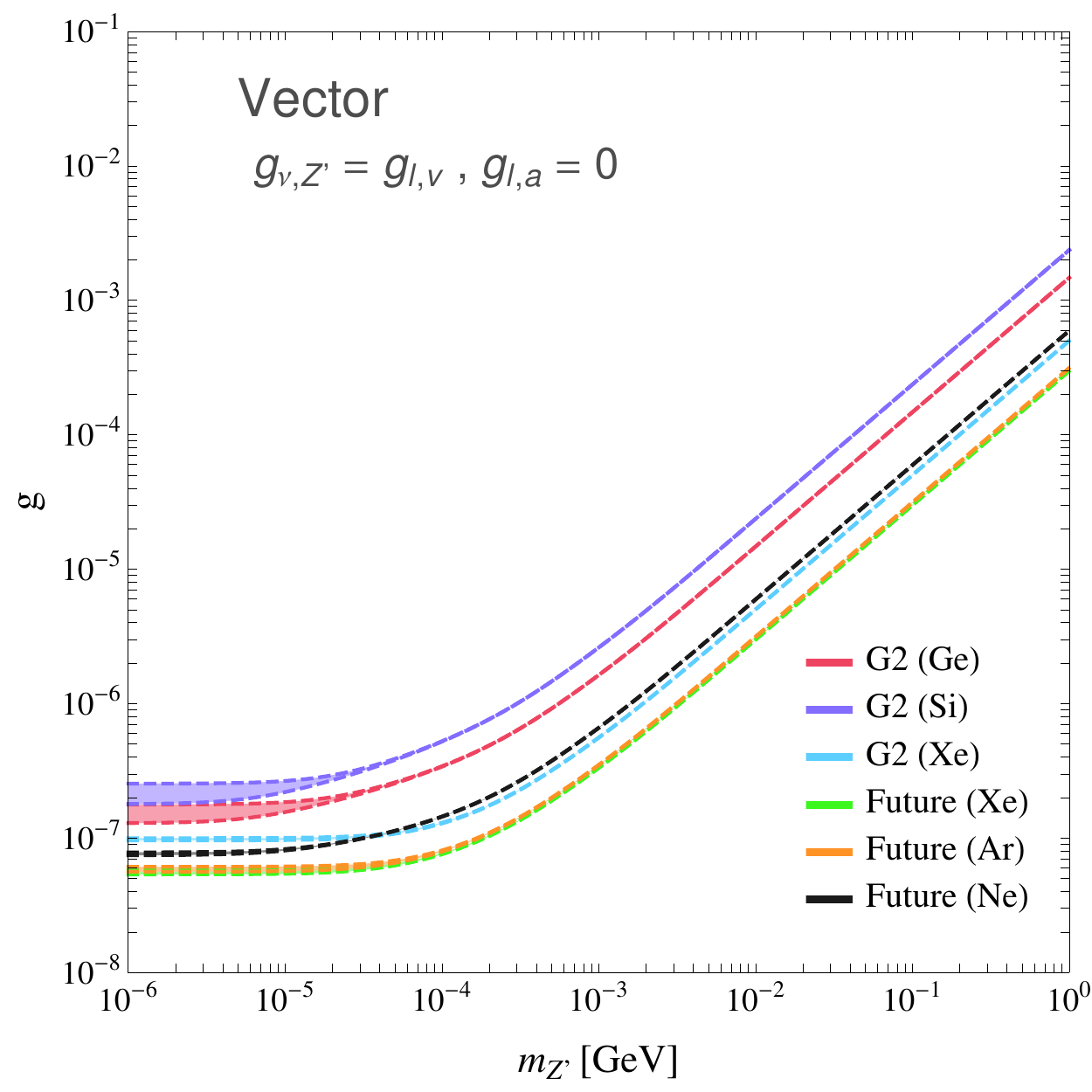} & \includegraphics[width=.45\textwidth]{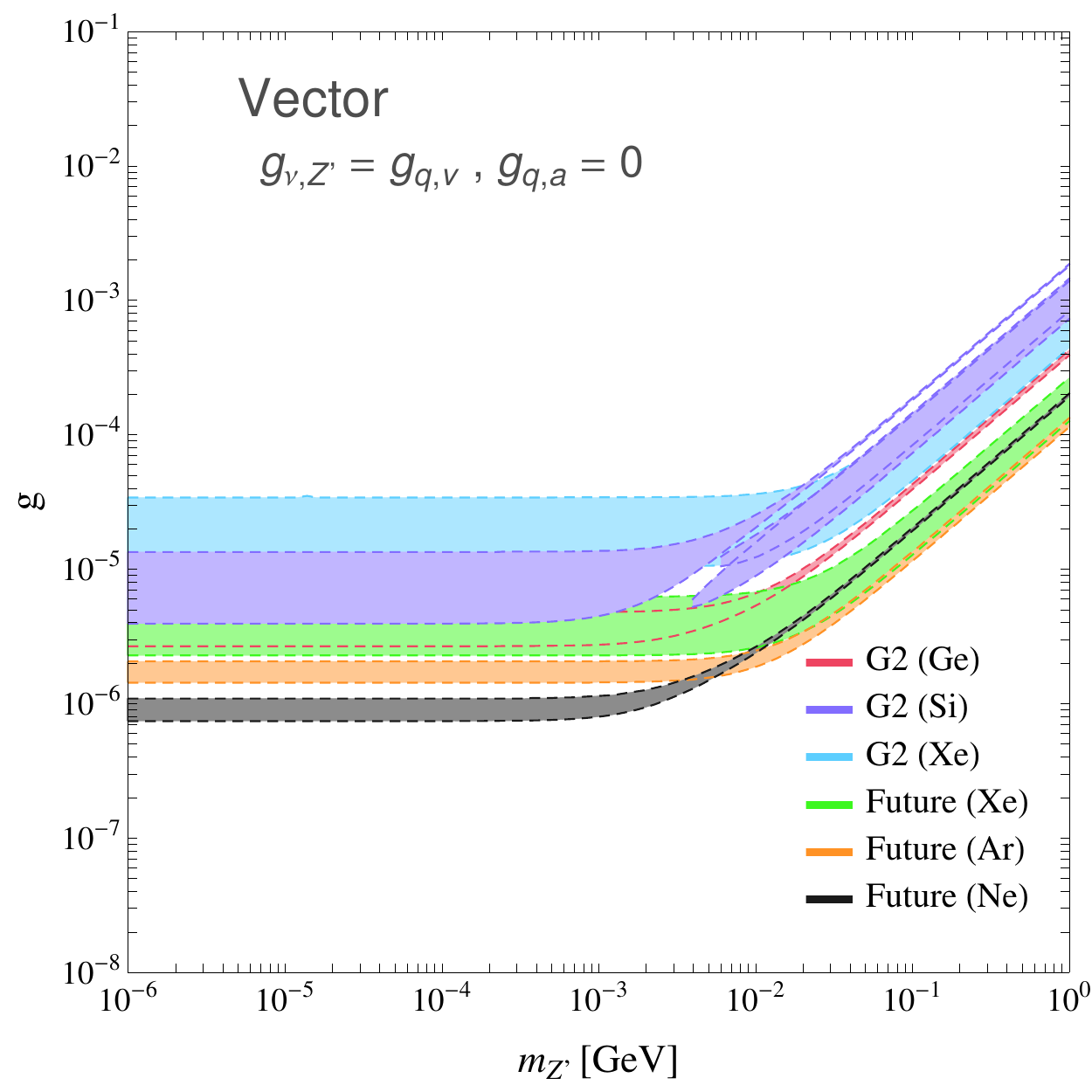}\\
\includegraphics[width=.45\textwidth]{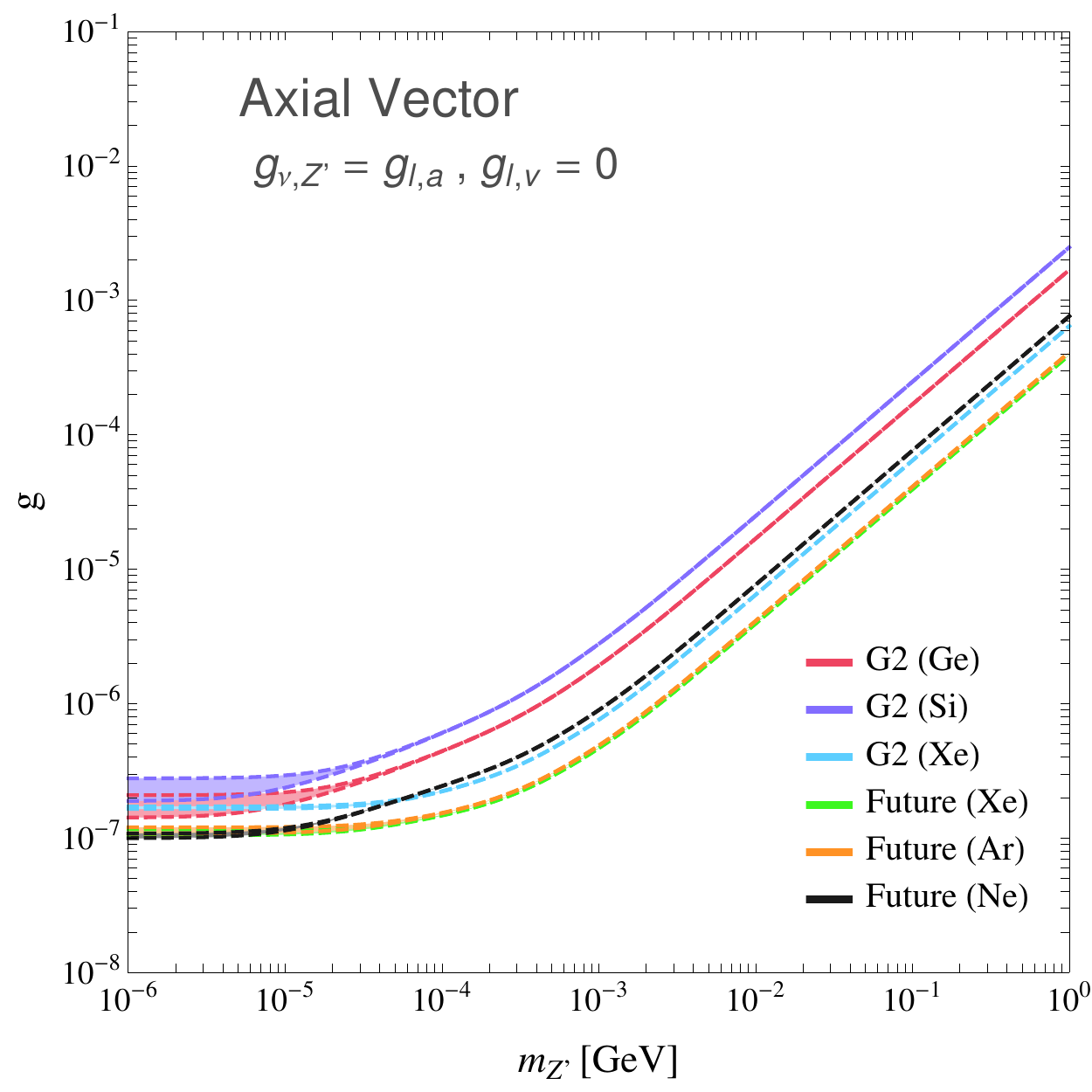} & \includegraphics[width=.45\textwidth]{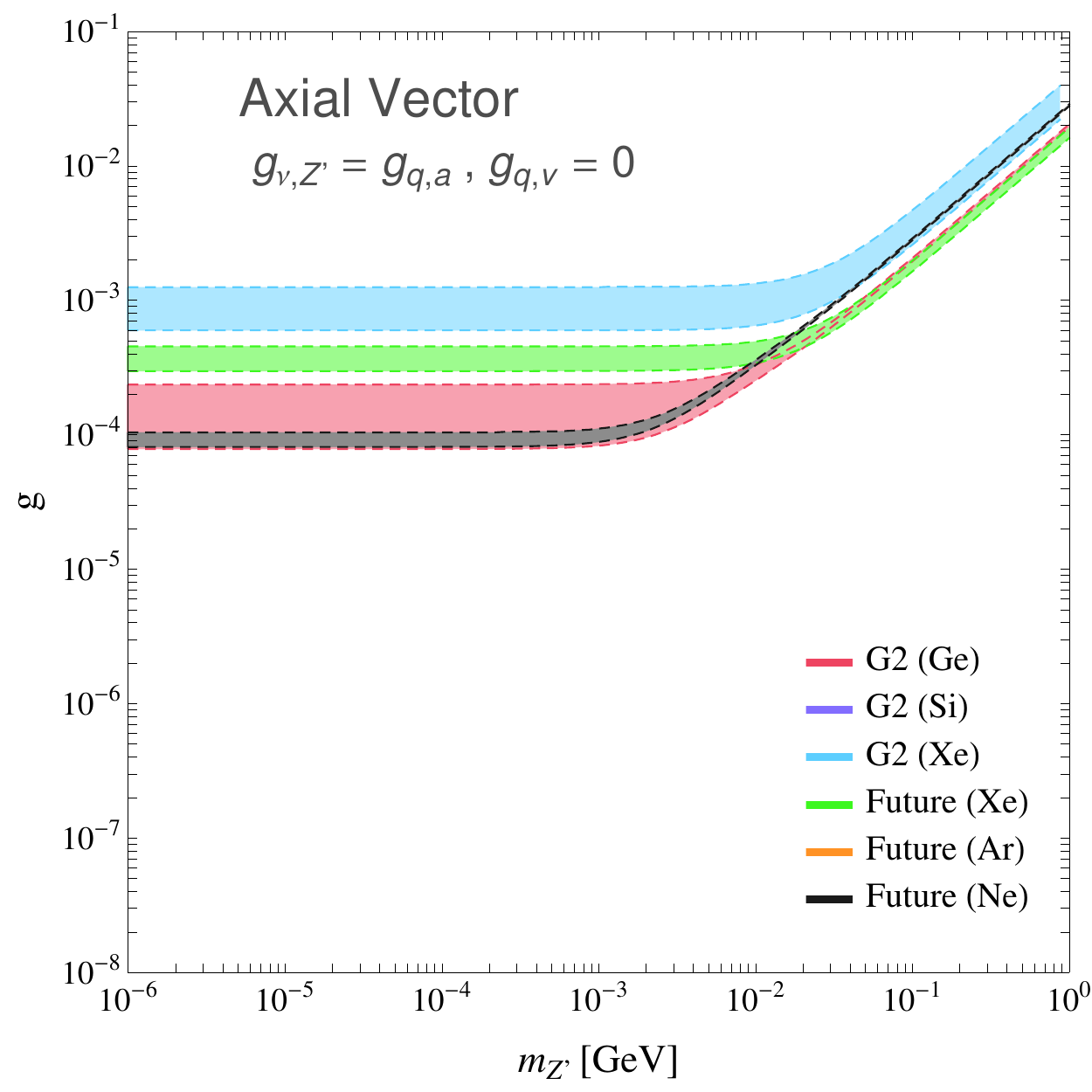}  
\end{tabular}
\caption{Electron recoil (left) and nuclear recoil (right) 90\% CL limits for a variety of target materials, using natural isotopic abundances. Top: scalar coupling; middle row: vector coupling; lower panels: axial vector coupling. The thickness of the bands represent the difference between the nominal (least constraining) and optimistic (most constraining) threshold configurations of Tab.~\ref{Table:exps}.
\label{newbosmass}}
\end{figure*}

\begin{figure*}[!htbp]
\begin{tabular}{c }
\includegraphics[width=.45\textwidth]{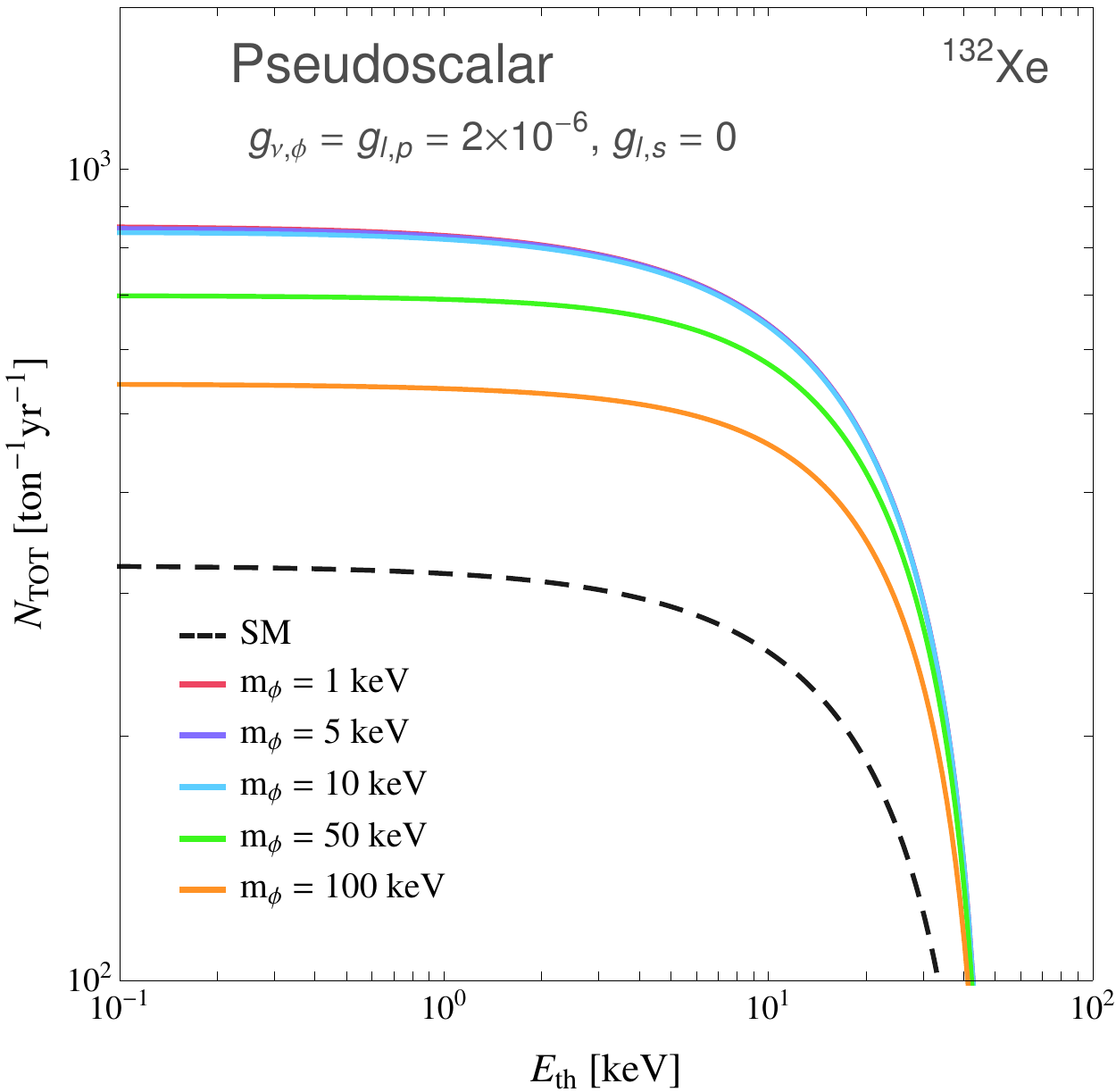}  \raisebox{-0.13cm}{\includegraphics[width=.455\textwidth]{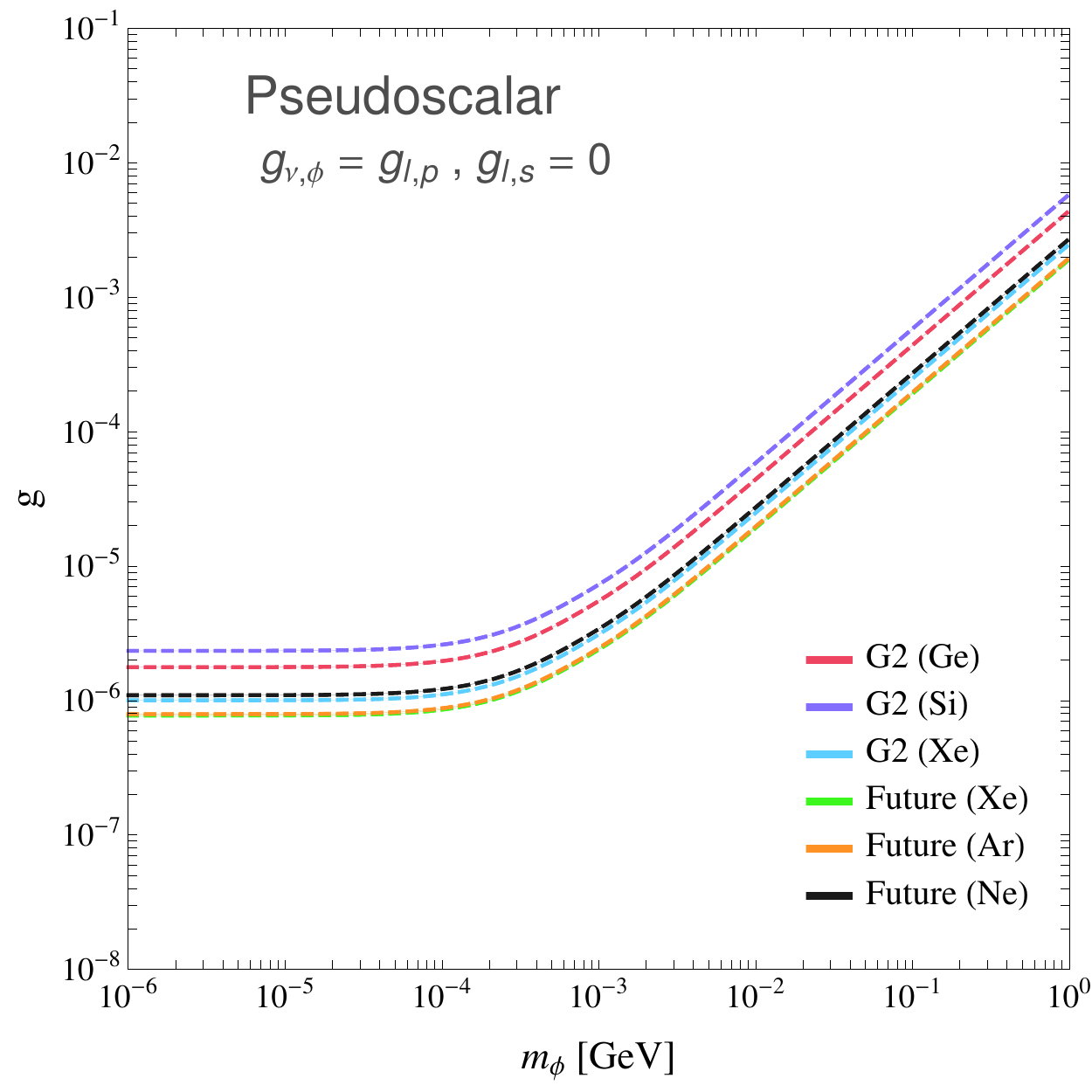}}
\end{tabular}
\caption{Electron recoil integrated rates (left) and sensitivity (right) for a pseudoscalar coupling.}
\label{fig:pseudoscalar}
\end{figure*}

The projected constraints on light scale physics are shown in
Figs.~\ref{newbosmass} and~\ref{fig:pseudoscalar}, for different mediators and target materials. The bands enclose the nominal and optimistic
scenarios defined in Tab.~\ref{Table:exps}. They are wider for
nuclear recoils (right panels) in comparison to electron recoils (left
panels) since the dependence with the threshold energy is more
pronounced.  
Depending on the mediator mass, electron recoils could
probe couplings below $10^{-6}$, while the bounds from nuclear recoils
would range from $10^{-3}$ to $10^{-6}$. In the case of a vector
mediator scattering off nuclei (middle right plot), the destructive interference with the SM $Z$ contribution may
lead to disconnected regions, e.g., for a G2 silicon-based
detector.
It is worth remembering at this point that we are basing our projections on the assumption that backgrounds can be removed. As discussed above, this is a reasonable hypothesis for the case of nuclear recoils but more challenging for electron recoils. 

\begin{figure*}
\includegraphics[width=1\textwidth]{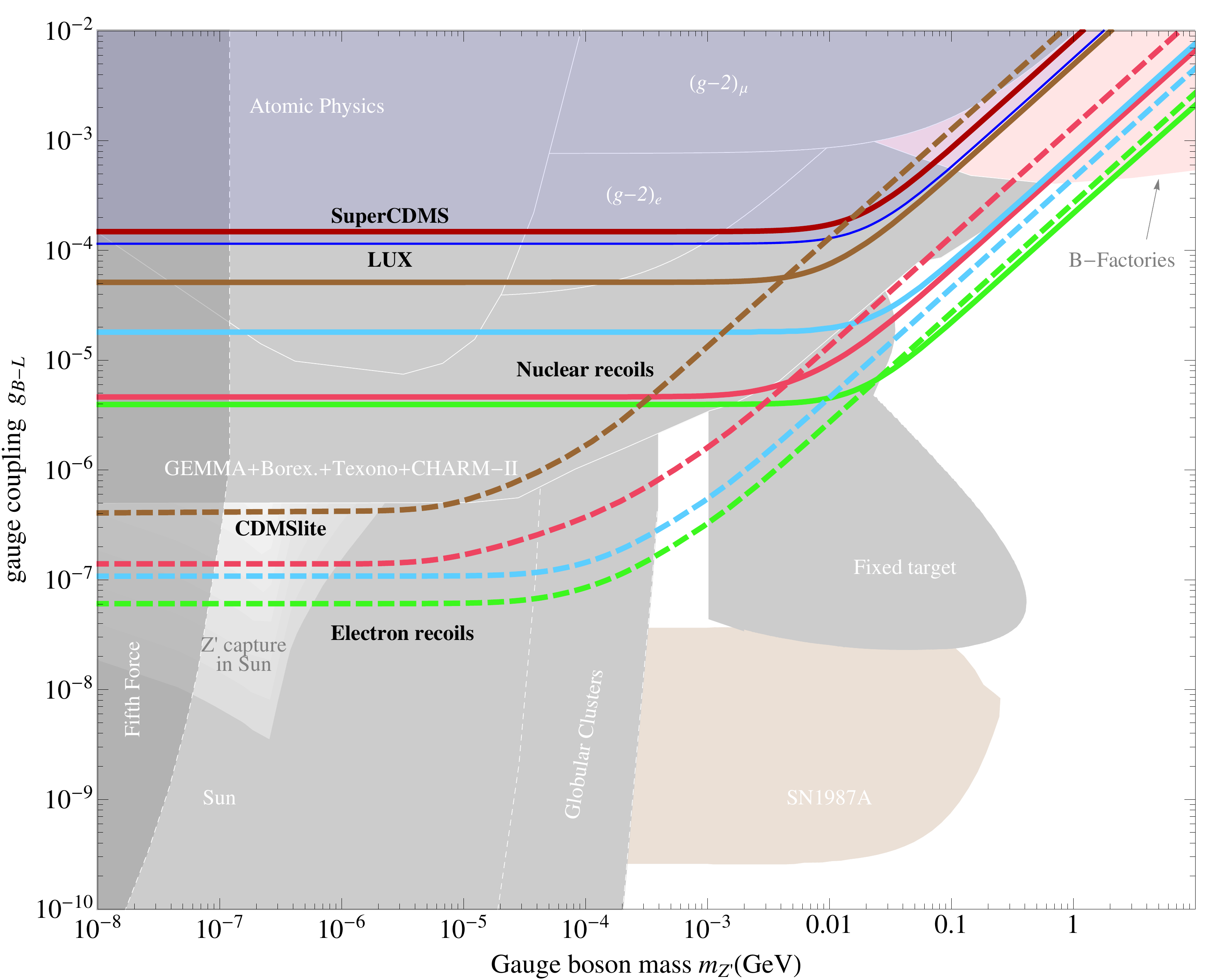} 
\caption{Projected 90\% CL constraints on the B-L model for nuclear recoils (solid lines) and electron recoils (dashed) in the optimistic scenarios for G2 germanium (red), G2 xenon (blue) and future xenon (green). We also show approximate bounds derived from the current SuperCDMS (red line), CDMSlite (brown solid and dashed lines), and LUX data (blue shaded region). Our bounds are overlaid on existing constraints. To translate these bounds to the other possible scenarios, one should keep in mind that some bounds (intermediate grey) only apply when the new mediator couples to leptons. The supernova bound (brown) only applies to couplings to baryons, while B-factory bounds (pink) require both. The fifth force constraint (dark grey) applies in either case. The grey regions, the neutrino scattering bound and the pink pregion, and the supernova limits are respectively taken from Refs.~\cite{Harnik:2012ni}, \cite{Bilmis:2015lja}, and \cite{Rrapaj:2015wgs}.}
\label{fig:BminusL}
\end{figure*}

\subsection{Bounds for a $U(1)_{B-L}$ model}

To put the sensitivity of future DD experiments in context, we
illustrate our results with the specific example of a light
$U(1)_{B-L}$ gauge boson, a construction that was studied in
Ref.~\cite{Harnik:2012ni} for $\nu-e$ scattering. In this case, a new
vector mediator couples to the $B-L$ quantum numbers of standard model
particles. Quarks therefore carry charge $1/3$ under this new
gauge coupling, while leptons have charge~$-1$. 

In Fig.~\ref{fig:BminusL} we present our bounds as before. The coloured lines are the result of this study. We use  the optimistic threshold scenarios of a G2 germanium (red lines) and xenon experiment (blue), as well as for a future DARWIN-like xenon target (green). We separate the limits that can be inferred from nuclear (solid lines) and electron recoils (dashed). As in the cases shown in Fig.~\ref{newbosmass}, electron bounds tend to do better, thanks to the larger $pp$ flux and to the closer kinematic matching between the solar neutrino energies and electron mass, allowing for higher recoil energies.

Our results in Fig.~\ref{fig:BminusL} are overlaid on excluded areas from previous studies, in the plane of gauge coupling $g_{B-L}$ versus mediator mass. A detailed description of each bound can be found in  Ref.~\cite{Harnik:2012ni} and references therein (see also Ref.~\cite{Bilmis:2015lja} for the TEXONO and CHARM-II limits). It should be emphasized that these limits are not model-independent, as they are sensitive to the coupling between the gauge boson and a specific fermion, as well as to the Lorentz structure of the coupling. These bounds fall into three broad categories:

\begin{itemize}
\item \textbf{Coupling to electrons (or muons) only}\\
 ``Atomic
physics'' (measurements of energy levels of atomic excited states),
``Sun'' and ``Globular Clusters'' (star cooling via the emission of
the mediator), ``Borexino'' (solar neutrinos scattering off
electrons), ``TEXONO'' and ``GEMMA'' (reactor neutrinos scattering off
electrons), as well as CHARM-II (accelerator neutrinos scattering off
electrons) all require a coupling to electrons. The region labeled as ``$Z'$ capture in Sun'' is not well understood:
although the Sun would not lose energy due to $Z'$ emission, solar
dynamics could be severely modified, and exact bounds have yet to be computed. The anomalous magnetic moment bounds require couplings to
electrons or muons. Moreover, these curves only apply to pure vector
couplings (e.g., the curve for axial vector couplings does not flatten at low
mediator masses~\cite{Studenikin:1998cs, Jegerlehner:2009ry}). 

\item \textbf{Coupling to electrons and/or quarks}\\
``Fixed target'' bounds require coupling to electrons only or both 
electrons and light quarks, depending if the experiment considered is 
an electron or proton beam dump.
For the first, the mediator is produced by radiation when $e^-$ collide 
with a target, while in proton dump experiments, the production is 
dominated by pseudoscalar meson decays (e.g. $\pi^0\to\gamma Z'$).
For both cases, the signature consists of $Z'$ decay to $e^+e^-$ (the
sharp cut on the left of this region corresponds to $2\,m_e$, below
which the production of two electrons is kinematically forbidden).
Notice that a larger coupling to neutrinos would enhance the mediator
invisible branching ratio, weakening this bound. The ``Fixed Target''
region shown in Fig.~\ref{fig:BminusL} includes only electron dump
experiments. Proton dump experiments are almost entirely within that region
and their inclusion will not change our conclusions.
The ``B-factories'' region requires non vanishing couplings to bottom
quarks.
\\
\item \textbf{Coupling to quarks only}\\
The ``SN1987A'' region (Supernova 1987A losing half of its energy via
$Z'$ emission) is sensitive to couplings to nucleons.  

\item \textbf{Any coupling}\\
Fifth
force searches (tests of gravitational, Casimir and van der Waals
forces) would be sensitive to all scenarios, except if the couplings
to protons, neutrons and electrons are proportional to the electric
charge, in such a way that the test bodies used in these experiments
are essentially neutral under the new interaction.
\end{itemize}

Finally, we present an estimate of the current bounds based on available data from the SuperCDMS Soudan  results~\cite{Agnese:2014aze}, CDMSlite~\cite{Agnese:2015nto}, and the latest LUX  data~\cite{Akerib:2015rjg}. For SuperCDMS we take an exposure of 577 kg days, and an energy range $E_{th}= 1.6$ keV, $E_{max}=10$ keV. We neglect detector resolution effects, and use the efficiency presented in Figure 1 of Ref.~\cite{Agnese:2014aze}. We take the 11 candidate events of this study as background. 
In the case of LUX, the exposure was 38.4 kg years, with an energy range $E_R = [1.1,25]$ keV. We take the efficiency from Figure 1 of Ref.~\cite{Akerib:2015rjg}. To model the detector resolution, we adopt a Gaussian smearing of $E_R$, with a width $\sigma_{E_R}/\mathrm{keV} \simeq 0.2 (E_R/\mathrm{keV})^{0.6}$ \cite{Akerib:2015wdi}.\footnote{This can be inferred via the relationship $E_R = 13.7\mathrm{\, eV} \cdot (n_e + n_\gamma)$, where $n_e$ and $n_\gamma$ are the measured electron and photon numbers, and the error on these quantities as a function of energy is shown in Figure 12 of \cite{Akerib:2015wdi}.}

In the case of CDMSlite, we analysed separately electron and nuclear recoils separately. The exposure of its second run~\cite{Agnese:2015nto} was $70$ kg days, with a threshold as low as $E_R=0.056$~keV for electron recoils ($E_R\sim 150$ eV for nuclear recoils). We consider the efficiency, energy bands and background (we assume a flat background) defined in Fig. 3 and Table I of Ref.\,\cite{Agnese:2015nto}.  We have derived independent constraints on nuclear recoils (solid brown line) and electron recoils (dashed brown line), since CDMSlite cannot distinguish between these.  In this case spectral information is available as the data are split into four energy bins, and we modify the likelihood ratio analysis accordingly.

The current and future bounds presented here make it clear that, in the case where a new mediator couples only to baryons or charged leptons, DD experiments actually lead to the strongest constraints for large regions of the parameter space. In the case of lepton-only coupling, the strong electron recoil limits push into the $\sim$ MeV mediator window, inaccessible to fixed target experiments, and can strengthen limits by a factor of $\sim 5$ for mediators above $\sim 100$ MeV. If the new mediator couples more strongly to light quarks than to electrons, nuclear bounds from DD experiments dominate the parameter space, as the light grey and pink regions of Fig. \ref{fig:BminusL} cease to apply. In fact, the limits that we have derived in this paper from current results of LUX, SuperCDMS, and CDMSlite represent the strongest bounds on this scenario to date.\footnote{Although having a larger $\nu-q$ scattering than $\nu-e$ may seem unnatural, a simple example of a gauge invariant theory with this property would be a broken $B-3L_3$ gauge symmetry with a light mediator.}

\section{Conclusions}
\label{sec:conclusions}

In this work we have investigated the potential of direct detection
(DD) dark matter experiments to use the flux of solar neutrinos to
improve our understanding of particle physics and of the Sun, as well
as to probe the existence of hypothetical new messenger particles.

The observation of neutrino-electron scattering in next generation DD
experiments would lead to an independent measurement of $\st$ at
unprecedented low energies which cannot be reached by dedicated
experiments. A 4.5\% precision can be obtained in this measurement
from next-generation (G2) experiments, and hypothetical future Xe and
Ar experiments could reduce this down to 1.4\%. Future dedicated
neutrino experiments such as HyperKamiokande and JUNO
\cite{An:2015jdp} will further constrain the solar neutrino flux
normalizations, allowing an even more precise DD-inferred measurement
of $\st$.

Data from future DD experiments will also help constrain measurements
of solar parameters. Most notably, the total $pp$ neutrino flux could
be measured up to 0.6\% precision (compared with the current
experimental precision of approximately 10\%, and a projected percent-level precision from SNO+ \cite{Andringa:2015tza}). Other observables
related to the solar composition -- namely the opacity and metallicity
-- could also be improved, albeit with the help of complementary
probes.

Crucially, these forecasts are for setups very similar to DD experiments that will become operational and begin data taking over the next few years. In contrast, planned dedicated neutrino experiments are still up to a decade away. 

We have studied the conditions under which DD experiments can be
sensitive to solar neutrinos from the CNO cycle. We observed that this
would require a light target, combined with an extremely low energy
threshold and good discrimination between electron and nuclear
recoils. Due to the small atomic mass of Ne, gaseous TPCs of this
material with an energy threshold of approximately $0.1$~keV could be
ideal (although no discrimination NR/ER has yet been achieved at such
low energies).

Finally, we have studied the contributions from new physics to neutrino interactions, concentrating on simplified models designed to include the effects of light mediators (scalar, pseudoscalar, vector, and axial vector) which couple to neutrinos and either quarks or electrons. Figures \ref{newbosmass} and \ref{fig:pseudoscalar} show constraints that can be placed on such new particles based on future experiments. Through the specific example of a $B-L$ gauge boson, we have shown in Fig. \ref{fig:BminusL} that direct detection experiments are already competitive with bounds from other origins.

Existing bounds from other works rely heavily on  a new boson coupled to neutrinos and charged leptons. However, we have shown that a coupling of a new boson to neutrino and quarks is already constrained by both SuperCDMS and LUX. 

Electron recoil measurements can furthermore explore regions of the parameter space hitherto inaccessible to other searches, such as the gap below $m_\phi = 2m_e$ between fixed target and stellar bounds.

The expected event rate from coherent neutrino scattering in the next generations of dark matter direct detection experiments is often presented as a nearly impenetrable barrier to the search for new physics. Apart from being the first signal of coherent neutrino scattering, we have demonstrated here that DD experiments' sensitivity to the solar neutrino fluxes presents manifold scientific opportunities, both in terms of precision measurements and for the exploration of new physics, well beyond the original scope of these instruments.

\acknowledgments
We thank Henrique Araujo, Peter Ballett, Enectal\'i Figueroa-Feliciano, Igor Irastorza, El\'ias L\'opez-Asamar, and Aldo Serenelli. PM thanks the Mainz Institute for Theoretical Physics and Fermilab, and both CB and PM thank
the Aspen Center for Physics (NSF grant PHY-1066293 and the Simons Foundation).  We thank James Dent for pointing out typos in Table IV of an earlier version of this work. MF is grateful for support from the IPPP in the form of an associateship.  We acknowledge support from the STFC, the partial support of the Centro de Excelencia Severo Ochoa under Grant No. SEV-2012-0249, funding from the European Research Council under the European Union's Horizon 2020 programs (ERC Grant Agreement no.648680 and RISE InvisiblesPlus 690575), the Consolider-Ingenio 2020 program under grant MultiDark CSD2009-00064, and the FP7 ITN INVISIBLES (PITN-GA-2011-289442). TJ was supported by Durham University.


\bibliography{coherence.bib}
\end{document}

%% file: jdefs.tex
\def\araa{ARA\&A}%
\def\apj{ApJ}%
\def\prd{Phys.\ Rev.\ D}%